\documentclass[10pt,journal,compsoc]{IEEEtran}
\usepackage{fancyhdr} 

\usepackage{ragged2e}
\usepackage{xspace}
\usepackage{cite}
\usepackage{url}
\usepackage[export]{adjustbox}
\usepackage{graphicx}  
\usepackage{epstopdf}
\usepackage{subfig}
\usepackage{multirow}
\usepackage{hyphenat}
\newcommand{\parahead}[1]{\vspace*{0.4ex plus 0.15ex minus 0.15ex}\noindent %
  {\bfseries #1.}}

\renewcommand{\paragraph}[1]{\vspace{2pt plus 0pt minus 2pt}\noindent{\bfseries #1}}

\usepackage{graphics}
\usepackage{caption}

\makeatletter  
\newif\if@restonecol  
\makeatother

\usepackage[linesnumbered,ruled,vlined]{algorithm2e}
\usepackage{algpseudocode}
\usepackage[table,xcdraw]{xcolor}
\usepackage{amsmath}  
\usepackage{color}
\usepackage{diagbox}
\usepackage{amssymb}

\newcommand{\ie}{\emph{i.e.},\xspace}

\newcommand{\eg}{\emph{e.g.},\xspace}

\newcommand{\etal}{\emph{et al.}\xspace}

\newcommand{\sysname}{{\sf AdaEvo}\xspace}
\newcommand{\sysnameposs}{{\sf AdaEvo's}\xspace}

\begin{document}

\title{\centering{\justifying{AdaEvo: Edge-Assisted Continuous and Timely \\ DNN Model Evolution for Mobile Devices}}}
\author{Lehao Wang, Zhiwen Yu, \textit{Senior Member, IEEE}, Haoyi Yu, Sicong Liu, Yaxiong Xie, Bin Guo, \textit{Senior Member, IEEE}, Yunxin Liu, \textit{Senior Member, IEEE}}

\author{Lehao Wang, Zhiwen Yu, \textit{Senior Member, IEEE}, Haoyi Yu, Sicong Liu, Yaxiong Xie, Bin Guo, \textit{Senior Member, IEEE}, Yunxin Liu, \textit{Senior Member, IEEE}
\IEEEcompsocitemizethanks{
\IEEEcompsocthanksitem Lehao Wang, Zhiwen Yu, Haoyi Yu, Sicong Liu and Bin Guo are with School of Computer Science at Northwestern Polytechnical University, Xi'an, Shaanxi, China. E-mail: 	\{lehaowang, haoe\}@mail.nwpu.edu.cn, \{ zhiwenyu, scliu\}@nwpu.edu.cn, guobin.keio@gmail.com
\IEEEcompsocthanksitem Yaxiong Xie is with the Department of Computer Science and Engineering at the University at Buffalo, SUNY, New York, USA. E-mail:  yaxiongx@buffalo.edu
\IEEEcompsocthanksitem Yunxin Liu is with the Institute for AI Industry Research (AIR), Tsinghua University, Beijing, China. E-mail: liuyunxin@air.tsinghua.edu.cn
}
\thanks{
This work was partially supported by the National Natural Science Foundation of China (No. 61960206008), the National Science Fund for Distinguished Young Scholars (No. 62025205), the National Natural Science Foundation of China (No.  62032020, 62102317).
}
\thanks{(Corresponding authors: Zhiwen Yu and Sicong Liu)}
}

\markboth{IEEE TRANSACTIONS ON MOBILE COMPUTING}%
{Shell \MakeLowercase{\textit{et al.}}: AdaEvo: Edge-Assisted Continuous and Timely DNN Model Evolution for Mobile Devices}
\IEEEtitleabstractindextext{%
\begin{abstract}

\justifying
Mobile video applications today have attracted significant attention.
Deep learning model (\eg deep neural network, DNN) compression is widely used to enable on-device inference for facilitating robust and private mobile video applications.
The compressed DNN, however, is vulnerable to the agnostic data drift of the live video captured from the dynamically changing mobile scenarios.
To combat the data drift, mobile ends rely on edge servers to continuously evolve and re-compress the DNN with freshly collected data.
We design a framework, \sysname, that efficiently supports the resource-limited edge server handling mobile DNN evolution tasks from multiple mobile ends.
The key goal of \sysname is to maximize the average quality of experience (QoE), \ie the proportion of high-quality DNN service time to the entire life cycle, for all mobile ends.
Specifically, it estimates the DNN accuracy drops at the mobile end without labels and performs a dedicated video frame sampling strategy to control the size of retraining data.
In addition, it balances the limited computing and memory resources on the edge server and the competition between asynchronous tasks initiated by different mobile users.
With an extensive evaluation of real-world videos from mobile scenarios and across four diverse mobile tasks, experimental results show that \sysname enables up to $34\%$ accuracy improvement and $32\%$ average QoE improvement.

\end{abstract}
}

\maketitle

\section{Introduction}
\label{sec:intro}

\IEEEPARstart{A}{} broad spectrum of mobile applications today requires real-time video stream analytics,
such as mobile VR/AR~\cite{bib:mmsys2019:shi}, autonomous human-following drones~\cite{bib:itoit2018:ke}, 
vision-based robot navigation~\cite{bib:icufn2018:kim}, and autonomous driving cars~\cite{bib:iccv2015:chen}.
Video stream analytics tasks such as object detection and classifications heavily rely on deep neural networks (DNN), 
\textit{e.g.}, Faster R-CNN\cite{bib:nips2015:Ren}, YoloV3\cite{bib:arXiv2018:Redmon}, and FCOS~\cite{bib:iccv2019:fcos}.
Offloading the DNN inference to the edge server or cloud is a common practice to accommodate the intensive computational resources DNNs require.
Nevertheless, this approach introduces significant latency, privacy concerns, and fails to meet the real-time inference demands of mobile applications.
To this end, on-device DNN inference at the mobile end has attracted significant attention due to its  robustness~\cite{bib:mobicom2018:fang, bib:osdi2018:hsieh}.
By processing video streams locally, it reduces latency and ensures user privacy.
Researchers have presented a variety of DNN  compression methods, (\textit{e.g.}, 
 quantization and pruning\cite{bib:iclr2016:han,bib:cvpr2019:wang,bib:nips2015:han,bib:iclr2016:li}, 
tensor decomposition\cite{bib:cvpr2019:kim,bib:nips2015:novikov} , knowledge distillation\cite{bib:arXiv2015:Hinton,bib:aaai2019:heo} and online compression\cite{bib:liu2021adaspring}) to facilitate the deployment of DNNs on resource-constrained mobile ends.

However, compressed DNN-based mobile video analytics inevitably have the data drift problem~\cite{bib:csur2014:survey,lu2018learning,bib:nsdi2022:Bhardwaj},
\textit{i.e.}, the live video stream captured by the mobile-end camera diverges from videos used for training, which leads to an \textit{accuracy drop} in real-world applications.
Essentially, the reasons are two-fold:
\textit{(i) Data distribution shift.} The data distribution shift characterizes the difference between the distributions of the training dataset and the testing data.
A DNN with enormous parameters is generalizable and will still be affected by data distribution shift because it violates the IID assumption~\cite{bib:quinonero2008dataset, bib:moreno2012unifying}. 
\textit{(ii) DNN compression.} 
The accuracy drop problem worsens for a compressed DNN because it cannot generalize well with the pruned structure and sparse parameters, resulting in considerable drops in accuracy across dynamic mobile scenarios.
For example, in autonomous driving applications, the data distribution of the freshly captured videos varies significantly because of the dynamically and frequently changing mobile scenes. 
The accuracy of the compressed DNN fluctuates dramatically, and the out-date DNNs may even be too low-quality.

In view of those challenges, various efforts have been explored~\cite{bib:nsdi2022:Bhardwaj, bib:ICLR2017:yoon2018, bib:iccv2019:Mullapudi} (see more discussions in \S \ref{sec:related}).
Among them, the edge server-assisted continuous model evolution is one of the most practical and promising solutions, where the DNN model is incrementally retrained at the \textit{edge server} with freshly captured video streams~\cite{bib:nsdi2022:Bhardwaj,bib:IITJ2022_Jia}.
The \textit{mobile end} sends an evolution request, and uploads recently recorded video frames to the edge server when the inference accuracy of the compressed DNN falls below a tolerable threshold.
The edge server tackles all the asynchronously arrived model evolution requests from multiple mobile ends.
The edge server and all mobile ends establish a \textit{holistic system} and thus become closely correlated.
Each action (\eg video stream sampling, DNN evolution, \etal) taken by any mobile/edge member in this system affects the system's overall performance, especially when the resources of mobile and edge devices are limited.
For example, the more video frames uploaded by the mobile end, the higher accuracy the retrained DNN gains, which, on the other hand, results in longer retraining time and more allocated edge resources for this mobile end.
And we face the following challenges on the mobile and edge sides, respectively.

First, it is non-trivial for the \textit{mobile end} to accurately and timely estimate the \textit{accuracy drop} of the deployed DNN for deciding when to trigger the evolution requests and how to sample the most suitable  \textit{video frames} for uploading.
Either the fixed evolving frequency~\cite{bib:iccv2021:Khani} or server-side accuracy assessment~\cite{bib:iccv2019:Mullapudi} leads to unnecessary model evolutions, increase the workload of the server side, and even lags some necessary evolution tasks.
And it is difficult to accurately predict the accuracy drop without data labels and with limited data storage space.
Moreover, it is challenging to select the most miniature set from live video streams that represent new scenarios and contribute to the accuracy gain of mobile DNN evolutions.
It is a trade-off between DNN evolution accuracy gain and efficiency.
Existing methods using predefined sampling rates~\cite{bib:iccv2021:Khani} cannot fully represent the new scenes if the sampling rate is low. 
And a higher sampling rate may result in lower evolution efficiency with long data uploading and model evolution delay.

Second, at the \textit{edge server side}, it is intractable to schedule all the asynchronously arrived model evolution requests 
and allocate memory and computing resources for each request separately for balancing the overall performance. 
Our insight is that the memory resources could easily become the \textit{bottleneck} with the increasing number of mobile ends this edge server serves, causing significantly enlarged model evolution latency. 
Without a timely evolved DNN, the mobile end has to rely on the outdated model and thus suffers from increased inference degradation. 
And one evolution task that takes a long time to finish will delay the execution of others at the resource-constrained edge server,
introducing a smaller overall proportion of high-quality service time to the entire DNN life cycle.
Allocating more memory or computing resources for one evolution task may increase the time another task waits before execution and the time the edge server takes to evolve other tasks.
%

Given the above challenges, this paper presents \sysname, a framework to fairly share the valuable edge resources among multiple mobile ends and simultaneously maximize the overall quality of all served mobile DNNs.
\textit{First, on the mobile end-side}, we design the detection confidence metric to measure the DNN accuracy drop.
And based on this metric, we detect the data drift start and end time to determine the evolution trigger time-point.
In addition, to sample minimal video frames that fully reflect the new mobile scenes at each evolution, we design adaptive frame selection strategies based on the detected diverse data drift types, thus maximizing the performance gains from model evolution.
\textit{Second, on edge server-side}, 
we build a profiler for every evolution task, including the required memory to run the evolution, 
the time the server takes to retrain the model, and the accuracy gain the mobile end will obtain with the evolved DNN. 
The online task scheduler consists of two steps, \ie task selection and resource allocation. 
We formulate the evolution task selection as a Tetris stacking problem and propose a dynamic programming-based algorithm.
We design the server resource allocation strategy to adaptively allocate the memory according to the request, maximize throughput, and minimize the average retraining time.

We implement \sysname system with PyTorch~\cite{bib:nips2019:Paszke} and Flask ~\cite{bib:Ronacher2015flask} frameworks over four real-world mobile video applications.
Experimental results show \sysname enables up to $34\%$ accuracy improvement and $32\%$ average QoE improvement. 
The main contributions of this paper can be summarized as the following three points:
\begin{itemize}
    \item 
    To the best of our knowledge, we are the first to systematically formulate the edge-assisted DNN evolution in a holistic mobile-edge system and optimize the overall proportion of high-quality DNN service time to the entire life cycle of multiple mobile DNNs.

    \item We estimate the DNN accuracy drop to determine the evolution trigger time-point and design the adaptive frame sampling strategies for different types of data drift on the mobile end side. 
    And we propose the dynamic programming-based evolution task selection algorithm to maximize the average quality of all mobile DNNs with limited edge resources.

    \item Experiments show that \sysname achieves the best accuracy gain and highest average quality for heterogeneous and dynamic mobile live videos compared to other baselines.
    Also, it adaptively schedules the edge server resources for balancing varying asynchronous DNN evolution requests. 
\end{itemize}

In the rest of this paper, we propose a system overview in \S \ref{sec:formulation}, elaborate functional modules in \S \ref{sec:design_1} and \S \ref{sec:edge}, show evaluations in \S \ref{sec:experiment}, present the motivation in \S \ref{s_motivation}, review related work in \S \ref{sec:related}, and conclude in \S \ref{sec:conclude}.



\section{Overview}
\label{sec:formulation}


This section presents an overview of \sysname.

%


\begin{figure*}
    \centering
    \includegraphics[width=0.54\linewidth]{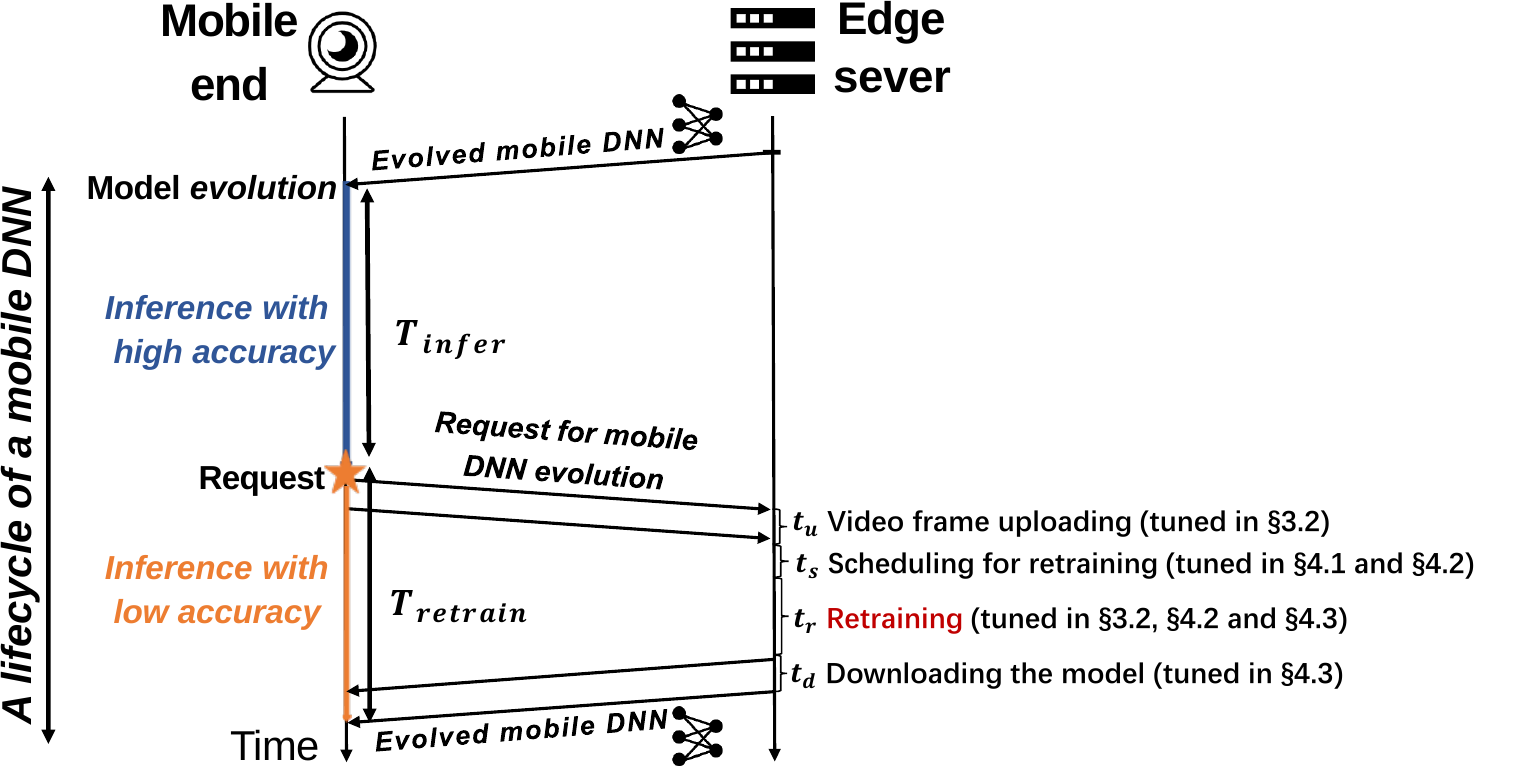}
    \caption{Illustration of four-time segmentation in the edge-assisted mobile DNN evolution workflow. 
    The inference time with robust DNN model $T_{infer}$, video uploading time $t_u$, and model retraining time $t_r$ are tunable in the mobile edge system.}
    \label{fig:ms_single}
\end{figure*}
    
%

\subsection{Problem Formulation}
\label{sec:form}
%
The fundamental goal of edge-assisted continuous DNN evolution is to maximize the \textit{Quality of Experience} (QoE) of the mobile application user.
The inference accuracy of the mobile DNN is the dominant factor that affects the mobile user's QoE, 
which, however, decreases with time because of the data drift problem we will expound in \S \ref{sec:314}.
Therefore, we formulate the \textbf{QoE of each mobile DNN $i$} as the proportion of high-quality DNN service time to the mobile DNN's full life cycle:
\begin{equation}
  Q_i = \mathcal{A}_i(t) \times \frac{T_{{infer}_i}} { T_{{infer}_i} + T_{{retrain}_i}}
\label{eq_q_single}
\end{equation}
where $\mathcal{A}_i(t)$ represents the time-varying inference accuracy of the deployed mobile DNN $i$.
The time $T_{{infer}_i}$ tells us how long the DNN works robustly with an accuracy greater than a threshold.
$T_{{retrain}_i}$ gives us the length of the period \textit{from} the mobile user initiating its evolution request for accuracy calibration \textit{to} the user finishing downloading the retrained model,
as shown in Figure~\ref{fig:ms_single}.
The total time $T_{infer}  + T_{retrain}$ represents the entire \textit{life cycle} of a particular mobile DNN.

To maximize $Q_i$, our key idea is to maximize the time ratio $R_t= T_{infer}/(T_{infer} + T_{retrain})$,
so that \textbf{a large portion of the DNN model's life cycle is spent on high\hyp{}quality inference}.
Maximizing the ratio requires maximizing $T_{infer}$ and minimizing $T_{retrain}$.
$T_{infer}$ depends on the accuracy decreasing speed of $A(t)$ and can be maximized if we obtain a generalizable model by evolution.
As shown in Figure \ref{fig:ms_single}, the time $T_{retrain}$ consists of four time segments:
\begin{equation}
    \quad T_{retrain} = t_u + t_s + t_r + t_d
   \label{eqn:t_retrain}
\end{equation}
where $t_u$ is the video frame uploading time,
$t_s$ is the scheduling time the edge server takes to schedule the request, 
$t_r$ is the model retraining time, and $t_d$ is the evolved model downloading time.
Therefore, \textbf{shortening any of the four-time segments contributes to the minimization of $T_{retrain}$}.
The time $t_u$ the user takes to upload video frames depends on the network capacity and the size of the video frames.
Reducing the number of video frames helps shorten  $t_u$ but would reduce the retraining data's quality and, thus, the retrained model's accuracy. Similarly, reducing the size of parameters for DNN evolution can reduce both retraining time $t_r$ and downloading time $t_d$.

%

We formulate the \textbf{average QoE of N mobile DNNs} as $Q_{ave} = \frac{1}{N}\sum^N_{i=1} Q_i$. 
Where $Q_i$ is the QoE of the $i$-th mobile end.
To reflect the different needs of users, we refine the elements in $Q_{ave}$ as $ Max \quad Q_{avg} = \frac{1}{N} \sum_{i=1}^N \lambda_i Q_{i} $.
$\lambda_i$ represents the model evolving urgency, and it is a function of the current inference accuracy $\mathcal{A}_i(t)$ and accuracy drop $\Delta \mathcal{A}$. We calculate $\lambda_{i}$ by $\lambda_{i} = \frac{100}{\pi}*[\arctan(\pi*(\frac{\Delta \mathcal{A}}{\mathcal{A}_i(t)}-0.8))+\frac{\pi}{2}]$ to limit the function value roughly within [0, 100], where $\mathcal{A}_i(t)$ can be represented by $CLC$.
And the edge server implements an evolution task scheduler to maximize the \textit{average QoE} of all the $N$ mobile DNNs it serves.

Purely maximizing the average QoE forces the task scheduler to give larger execution opportunities and allocate more GPU resources to tasks that contribute more to the QoE, causing fairness problems among tasks.
Therefore, we add two penalty terms to the average QoE to favor those tasks that have been waiting for a long time in the task queue, \ie
$Q_t = Q_{avg} - \mathcal{SD}_{i=1}^{N}t_{i_s} -  \mathcal{SD}_{i=1}^{N}t_{i_r}$.
Where $\mathcal{SD}_{i=1}^{N}t_{i_s}$ and $\mathcal{SD}_{i=1}^{N}t_{i_r}$ is the mean square error of $N$ request's scheduling time $t_{s}$ and retraining time $t_r$, respectively. 
We formulate the optimization problem as follows.
\begin{eqnarray}\label{equ_opt}
\underset{\varphi_i\in\{0,1\}}{\arg\max} \quad  Q_t &=&
Q_{avg} - \mathcal{SD}_{i=1}^{N}t_{i_s} -  \mathcal{SD}_{i=1}^{N}t_{i_r}
\nonumber \\
s.t. &\qquad& \sum_{i=1}^N{\mathcal{M}_i}\leq{\mathcal{M}_s} ,
\sum_{i=1}^N{\mathcal{C}_i}\leq{\mathcal{C}_s} 
\label{equ:opt}
\end{eqnarray}
where $\mathcal{M}_i$ and $\mathcal{C}_i$ are the allocated memory and computing resource for task $i$.
$\mathcal{M}_{s}$ and $\mathcal{C}_{s}$ are the dynamically available memory and computing resources of the edge server.
The task scheduler adjusts the scheduling results when a new task arrives or a running task finishes. 
Here, we face two challenges: \textit{(i)} Due to different model evolving urgency and limited edge resources, there is a trade-off between reducing the ${t_{i_s} + t_{i_r}}$ of the specific tasks and all tasks.
\textit{(ii)}
It is NP-hard (reduction from the Knapsack problem~\cite{bib:knapsack_problem}) to dynamically select the appropriate task combination from all asynchronous requests with different resource demands to fit them into the GPU with dynamic resource availability and maximize the average QoE.

\begin{figure*}[t]
    \centering
    \includegraphics[width=.86\textwidth]{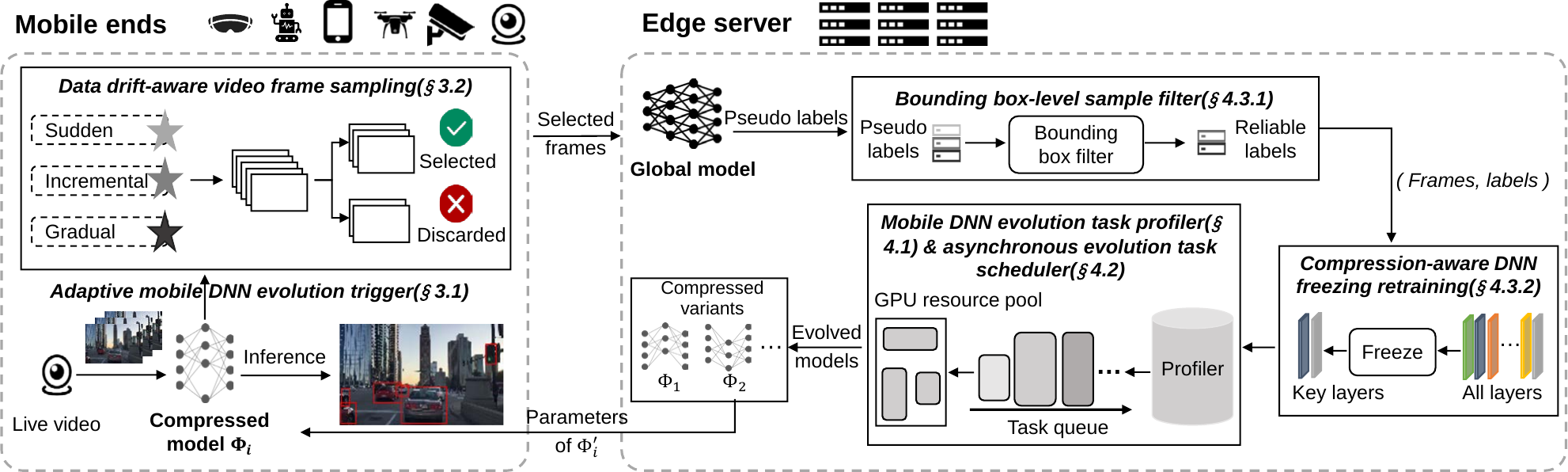}
    \caption{Illustration of the edge-assisted mobile DNN evolution system pipeline.}
    \vspace{-3mm}
    \label{fig:design_1}
\end{figure*}

\subsection{Characteristics of Mobile Data Drift}
\label{sec:314}  
In agnostic and changeable mobile scenes, live video data distribution shifts due to the influence of weather, lighting, or other factors, producing the phenomenon of data drift.
Specifically, given a segment of data in the time interval $[0,t]$, $V_{0,t} = \left \{(X_0, y_0), (X_1, y_1),..., (X_t, y_t)\right \}$, where $(X_i, y_i)$ denotes a sample at the moment $i$, $X_i$ is the feature vector, and $y$ is the label.
We define $P_t(X,y)$ as the joint probability distribution of $V_{0,t}$ and consider the data drift as the change in the joint probability of the data at the moment $t$, \ie $\exists t: P_t(X,y) \ne P_{t+1}(X,y)$.
The changes in data distribution can lead to a decrease in inference accuracy, \ie a model that performs well in historical scenes is not guaranteed to obtain the same performance level in new scenarios.

Moreover, the mobile data constantly changes in various patterns (we defer more details to \S~\ref{sec:43}).
To accurately identify the data drift for understanding the changes in mobile data distribution and designing targeted evolution methods, 
we classify the mobile data drift into three types, \ie \textit{sudden drift, gradual drift, and incremental drift}, shown in Figure~\ref{fig:drift_type}. 
Different types tend to indicate different accuracy drop patterns in live videos.
Specifically, \textit{sudden drift} refers to replacing old data distribution with the new one immediately.
For \textit{gradual} and \textit{incremental} drift, it will take a more extended period to complete the transition from the old data distribution to the new one.

\subsection{System Overview}
\label{sec:sys_overview}
To solve the above problems, \sysname presents the following components in both the mobile-end and edge-server sides.

\textbf{\textit{(i) Mobile end}}.
Each video frame captured by the mobile end goes into three components:
the \textit{compressed DNN} for video inference, 
the \textit{adaptive mobile DNN evolution trigger} module for evolution time determining, and
the \textit{data drift-aware video frame sampling} module for elite data uploading, as shown in Figure~\ref{fig:design_1}.
The \textit{adaptive mobile DNN evolution trigger} module quantitatively predicts the accuracy drop on live videos to judge whether data drift occurs and determine the evolution trigger time-point (\S\ref{sec:42}).
To balance the size of uploaded video frames and the accuracy gain of the evolved DNN, 
the \textit{data drift-aware video frame sampling} module identifies the current data drift type and selects the suitable frames which contribute the most to the edge server (\S\ref{sec:43}).

\textbf{\textit{(ii) Edge server}}.
The edge server retrains the specific mobile DNN once receiving video frames uploaded by the mobile end.
The edge server implements a \textit{bounding box-level sample filter} module (\S\ref{s:label_gen}) 
to automatically generate labels using a global model together with the uploaded frames.
The model retraining time $t_r$ is bound up with the size of the retrain data and the number of DNN parameters the edge server needs to evolve.
To shorten $t_r$, the edge server adopts \textit{compression-aware DNN freezing retraining} module (\S\ref{subsec:train_compress}) 
to freeze parameters that have negligible impact on retraining accuracy calibration.
The server specializes and retrains DNN according to the hardware specifications of the mobile end, as shown in Figure~\ref{fig:design_1}.
Reducing the retrained parameter size also helps decrease the downloading time $t_d$. 

The scheduling time $t_s$ equals zero if only one mobile end uses the resources of the edge server.
And the $t_s$ is tunable by the task scheduling with multiple mobile ends sharing the server's resources.
We present the \textit{ asynchronous evolution task scheduler} (\S \ref{s_task_scheduler}) for the edge server to resolve the above intractable problem. 
The scheduler employs a dynamic programming-based evolution task selection algorithm to select the optimal task subset from to-be-scheduled tasks for achieving the maximized overall QoE $Q_{ave}$ meanwhile satisfying the server's dynamic GPU resource constraint $\mathcal{M}_s$ and $\mathcal{C}_{s}$.
To estimate each task's GPU memory demand $\mathcal{M}_i$, 
evolution accuracy gain $\Delta \mathcal{A}$, and retraining time $t_{i_r}$ before scheduling, we also present a timely and accurate \textit{mobile DNN evolution task profiler} in \S \ref{s_estimator}.
Specifically, we calculate the memory demands by summing the memory resources occupied by the DNN size, retraining data, and the metadata generated during the retraining.
We estimate the accuracy gain by fitting a training epoch-accuracy curve with the training progress.
And, we predict the retraining time using a lightweight neural network trained by $200$ samples of diverse retraining settings and the corresponding retraining time.

\begin{figure}[tb]
    \centering
    \includegraphics[width=0.54\linewidth]{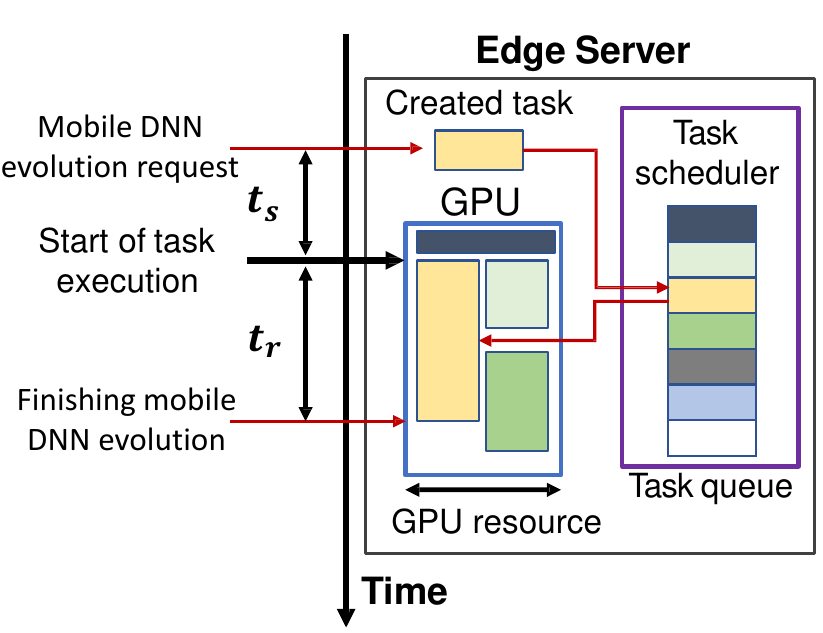}
    \caption{Illustration of the edge server handling DNN evolution requests from multiple asynchronous mobile ends. 
    The task scheduling and retraining delay $t_s$ and $t_r$ are tunable at the edge server side.}
    \label{fig:ms_multi}
    \vspace{-3mm}
\end{figure}

\textbf{\textit{(iii) Overall pipeline}}.
Diverse mobile ends (\eg vehicles and robots) load different compressed DNNs, which we regard as mobile DNNs, to meet specific application requirements and mobile platform constraints.
Over time, each mobile end needs to initiate the mobile DNN evolution request independently once it is in a low inference accuracy state on the live video.
And we offload the computation-intensive DNN evolution task to the edge server.
The edge server immediately creates a corresponding evolution task after receiving one request from the mobile end and then puts the created task into a task queue, as shown in Figure~\ref{fig:ms_multi}.
The edge server implements a task scheduler to arrange the execution of multiple evolution tasks.
According to the available resources, the task scheduler at the edge server selects appropriate tasks and puts them into the edge GPU for service.
As a result, each task needs to wait for a $t_s$ before retraining, and the task takes $t_r$ to finish.
After finishing one mobile DNN evolution task, the edge server generates one evolved DNN model and delivers this model to the mobile end.
The mobile end downloads the up-to-date DNN from the edge server and uses it for inference until the inference accuracy of the network falls below a threshold, as shown in Figure~\ref{fig:ms_single}.
To trigger the DNN evolution on the server side, the mobile end sends a request with a group of carefully selected video frames to the edge server.
The edge server reacts to the evolution request by loading the compressed DNN and the uploaded video frames for model retraining.
The edge server then notifies the mobile end about the completion of the retraining so it can download the evolved parameters immediately.
As a separate note, the evolved DNN specializes in diverse mobile application demands and dynamic mobile resource availability.

\section{\sysnameposs Mobile-end Design}
\label{sec:design_1}

This section elaborates on the \sysnameposs mobile-end design.

\subsection{Adaptive Mobile DNN Evolution Trigger}
\label{sec:42}

\sysname proposes an adaptive mobile DNN evolution trigger mechanism to reduce the number of DNN evolutions, thereby saving the resources of edge server and mobile ends and overhead for the system.
%
This section details metrics that measure the accuracy drop and trigger evolution.

\subsubsection{Accuracy Drop Prediction}
\label{metric:accuracy}
It is challenging to measure the accuracy drop exactly in dynamic mobile scenes with unlabeled data only relying on mobile ends.
Specifically, manual labeling is expensive and impractical.
And we can hardly accurately predict the accuracy drop only by the classification confidence~\cite{deng2021labels, boillet2022confidence}, which we can extract directly from the output of the softmax layer of the model classifier~\cite{deng2021labels} during model real-time inference. To this end, we estimate the real-time accuracy of the object detection DNN deployed on mobile ends by detection confidence $CLC = CC \times LC$, where $CC$ and $LC$ present the classification confidence and localization confidence, respectively.
As we will show in \S \ref{exp:211}, we have experimentally found that $CLC$ can accurately reflect the DNN accuracy drop caused by data drift, while $CC$ or $LC$ cannot.
We consider the prediction of localization confidence as a regression problem and employ a neural network consisting of two fully connected layers which only occupies a small number of resources for prediction.
Moreover, to train the neural network, we utilize a telescoping transformation of the ground-truth bounding boxes in the public dataset COCO~\cite{bib:Springer2014:Lin}, resulting in several candidate bounding boxes.
Then, we calculate the intersection-over-union (IoU) between each candidate bounding box and the corresponding ground-truth bounding box. 
And we combine the feature vector and IoU to train the regression model, which can be used to perform predictions of localization confidence immediately. 

\begin{figure}[tbp]
    \centering
    \includegraphics[width=.36\textwidth]{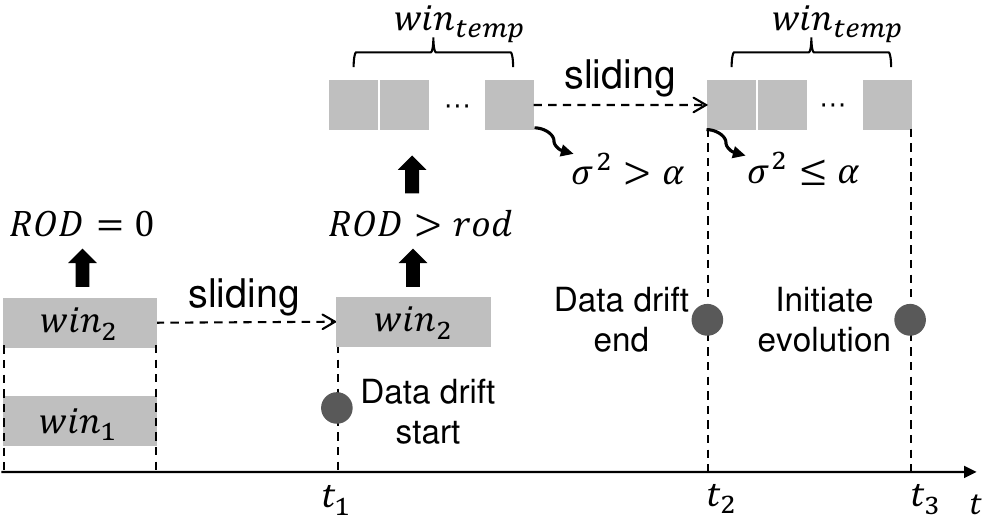}
    \caption{Detection of DNN evolution trigger time-point.}
    \label{fig:tup}
\end{figure}

\begin{figure*}[tb]
\centering
\subfloat[Sudden drift]{  
  \includegraphics[width=0.2\linewidth]{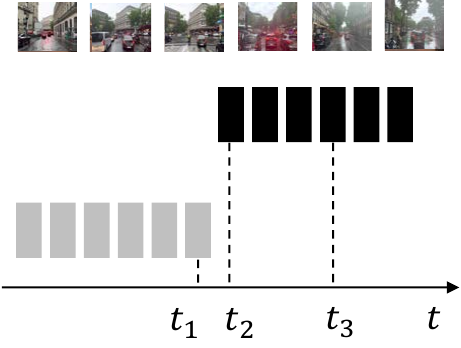}}
\hfill
\subfloat[Incremental drift]{
  \includegraphics[width=0.2\linewidth]{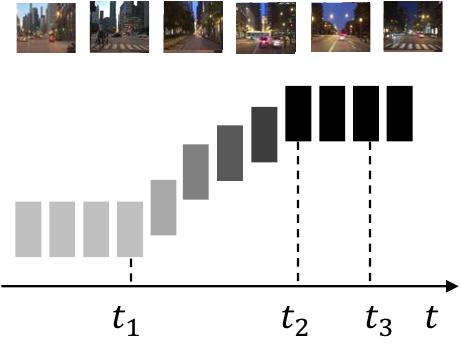}}
\hfill
\subfloat[Gradual drift]{
  \includegraphics[width=0.2\linewidth]{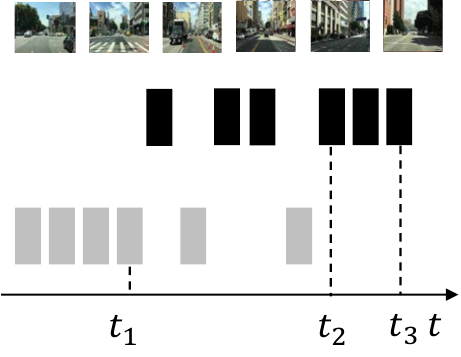}}
\hfill
\subfloat[Comparison result]{
  \includegraphics[width=0.36\linewidth]{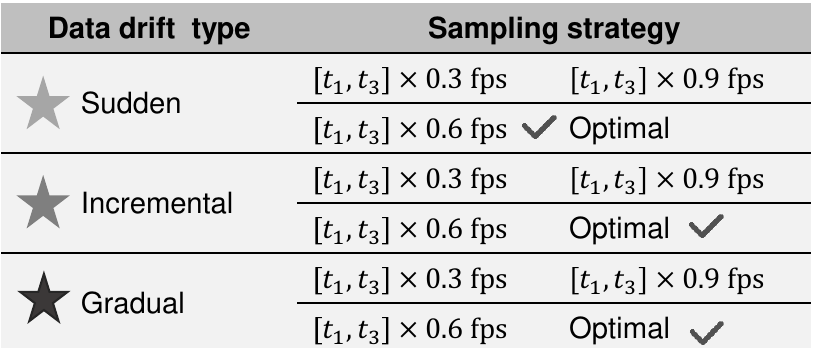}}
\hfill
\caption{Different types of data drift and the result of the fixed sampling rate and \sysname on these drift types.}
\label{fig:drift_type}
\end{figure*}

\begin{table*}[t]
\centering
\scriptsize
\caption{Performance comparison of the retrained DNN by using different video frames from the data drift process.}
\begin{tabular}{|c|c|c|cccccc|}
\hline
\multirow{2}{*}{\textbf{Mobile scenes}} & \multirow{2}{*}{\textbf{Performance metrics}}       & \textbf{w/o retraining} & \multicolumn{6}{c|}{\textbf{DNN evolution using different live video clips (\ie before, during, after data drift).}}                                                                                                                               \\ \cline{3-9} 
                               &                               &                & \multicolumn{1}{c|}{\textbf{before}} & \multicolumn{1}{c|}{\textbf{during}} & \multicolumn{1}{c|}{\textbf{before+during}} & \multicolumn{1}{c|}{\textbf{after}} & \multicolumn{1}{c|}{\textbf{during+after}} & \textbf{before+during+after} \\ \hline
\multirow{3}{*}{\textbf{cityA}}         & \textbf{mAP(IoU=0.50)}               & 0.484          & \multicolumn{1}{c|}{0.509}  & \multicolumn{1}{c|}{0.575}  & \multicolumn{1}{c|}{0.567}         & \multicolumn{1}{c|}{0.645} & \multicolumn{1}{c|}{0.693}        & 0.597               \\ \cline{2-9} 
                               & $\mathbf{T_{infer}}$~\textbf{(s)} & -              & \multicolumn{1}{c|}{43}     & \multicolumn{1}{c|}{188}    & \multicolumn{1}{c|}{206}           & \multicolumn{1}{c|}{420}   & \multicolumn{1}{c|}{526}          & 327                 \\ \cline{2-9} 
                               & $\mathbf{t_r}$~\textbf{(s)}           & -              & \multicolumn{1}{c|}{48}     & \multicolumn{1}{c|}{51}     & \multicolumn{1}{c|}{65}            & \multicolumn{1}{c|}{50}    & \multicolumn{1}{c|}{66}           & 94                  \\ \hline
\multirow{3}{*}{\textbf{cityB}}         & \textbf{mAP(IoU=0.50)}                    & 0.429          & \multicolumn{1}{c|}{0.469}  & \multicolumn{1}{c|}{0.537}  & \multicolumn{1}{c|}{0.552}         & \multicolumn{1}{c|}{0.638} & \multicolumn{1}{c|}{0.624}        & 0.56                \\ \cline{2-9} 
                               & $\mathbf{T_{infer}}$~\textbf{(s)} & -              & \multicolumn{1}{c|}{48}     & \multicolumn{1}{c|}{205}    & \multicolumn{1}{c|}{268}           & \multicolumn{1}{c|}{385}   & \multicolumn{1}{c|}{516}          & 341                 \\ \cline{2-9} 
                               & $\mathbf{t_r}$~\textbf{(s)}           & -              & \multicolumn{1}{c|}{45}     & \multicolumn{1}{c|}{47}     & \multicolumn{1}{c|}{63}            & \multicolumn{1}{c|}{45}    & \multicolumn{1}{c|}{65}           & 90                  \\ \hline
\multirow{3}{*}{\textbf{cityC}}         & \textbf{mAP(IoU=0.50)}                  & 0.397          & \multicolumn{1}{c|}{0.448}  & \multicolumn{1}{c|}{0.554}  & \multicolumn{1}{c|}{0.547}         & \multicolumn{1}{c|}{0.585} & \multicolumn{1}{c|}{0.603}        & 0.567               \\ \cline{2-9} 
                               & $\mathbf{T_{infer}}$~\textbf{(s)} & -              & \multicolumn{1}{c|}{51}     & \multicolumn{1}{c|}{177}    & \multicolumn{1}{c|}{215}           & \multicolumn{1}{c|}{326}   & \multicolumn{1}{c|}{487}          & 236                 \\ \cline{2-9} 
                               & $\mathbf{t_r}$~\textbf{(s)}           & -              & \multicolumn{1}{c|}{47}     & \multicolumn{1}{c|}{51}     & \multicolumn{1}{c|}{66}            & \multicolumn{1}{c|}{50}    & \multicolumn{1}{c|}{63}           & 87                  \\ \hline
\end{tabular}
\label{tb:videoSeg}
\end{table*}

\subsubsection{Detection of Evolution Trigger Time-point}\label{sec:422}

%
Our insight is that the onset of the accuracy drop is not always the optimal trigger time point for mobile DNN evolution.
Existing work~\cite{bib:iccv2019:Mullapudi} typically initiates an evolution request once it detects the accuracy drop, which leads to the lack of data from new scenes, frequent evolution, and shorter $T_{infer}$ duration that high-quality mobile DNN works.
Either the fixed or adaptive evolution frequency according to the historical contents will miss the best evolution time-point or causes costly frequent evolutions~\cite {bib:iccv2019:Mullapudi,bib:iccv2021:Khani}.

To detect the start and end of agnostic data drift, we leverage two sliding windows $win_1$ and $win_2$ to continuously track the accuracy drop by detecting ${CLC}_1$ and ${CLC}_2$, as shown in Figure~\ref{fig:tup}.
Initially, two windows are at the same position (\ie ${CLC}_1 = {CLC}_2$).
Then, we fix $win_1$ and slide $win_2$ over time.
Due to the data drift, the DNN works poorly in new mobile scenes, which causes ${CLC}_2$ to decrease.
We define the accuracy drop rate $ROD$ to reflect the degree that the model is affected by data drift.
\begin{equation}
ROD = \frac{{CLC}_1 - {CLC}_2}{{CLC}_1}
\end{equation} 
When $ROD$ is greater than the threshold $rod$, $win_2$ stops sliding.
At this time, the old data distribution starts to convert to the new data distribution, and we note the time as ${t}_{1}$.
Then, we set a temp sliding window $win_{temp}$ after ${t}_{1}$, which is divided into $n$ sub-windows. 
We slide $win_{temp}$ and keep calculating the variance $\sigma^2$ of the detection confidence of the n sub-windows.
When $\sigma^2$ is less than the threshold $\alpha$, we consider that the new data distribution is leveling off.
Meanwhile, the left border of $win_{temp}$ is recorded as ${t}_{2}$, indicating the end of the process of data drift.
And the right border of $win_{temp}$ is regard as ${t}_{3}$.

The time-point $t_3$ is optimal to trigger an evolution for diverse mobile data drift because it can ensure sufficient data of the new scene  and achieve the best overall performance in terms of accuracy, inference duration, and retraining time.
Specifically, we divide the continuous data drift phase into three states: before (\ie $\leq t_2$), during ($t_2 \sim t_3$), and after (\ie $\geq t_3$) the data drift, as shown in Figure \ref{fig:drift_type}.
For three different typical mobile data drift (\ie sudden, incremental, and gradual drift), Table~\ref{tb:videoSeg} shows the performance comparison of DNN evolution from the sampled data in the above three different phases.
We find that the model performs best with the retraining data during and after the data drift (\ie $t_2 \sim t_3 + \leq t_3$).
Although pending the data collected before the drift ($\ie \geq t_2$) can slightly generate a more significant accuracy gain, data expansion will increase the $t_r$.
Training with only the data after drift (\ie $\leq t_2$) usually leads to overfitting to new scenarios, reducing its generalization and thus shortening $T_{infer}$.
Therefore, \sysname adopts ${t}_{3}$ as the evolution trigger time-point and samples video frames from the data of the transformation process and the new scenario to balance the trade-off between the accuracy and retraining time, thereby obtaining better QoE of the single mobile end.

\subsection{Data Drift-aware Video Frame Sampling}
\label{sec:43}

To realize rapid DNN evolution with minimized video upload delay $t_{u}$, retraining delay $t_{r}$, and maximized accuracy gain,
\sysname proposes the fine-grained data drift-aware video frame sampling strategy.
Actually, in real-world scenarios, mobile ends encounter time-variant data drift types~\cite{lu2018learning} and uneven distribution of representative video frames.
The fixed sampling rate strategy (\eg 30~fps to 5~fps) cannot always select which frames reflect the characteristics of the new scene and thus can only obtain sub\hyp{}optimal performance~\cite{bib:iccv2019:Mullapudi,bib:iccv2021:Khani}.
Especially, for incremental and gradual drift, even the best-fixed sampling rate (\eg 0.6~fps) still loses some data important for DNN evolution.

Figure~\ref{fig:drift_type} (d) summarizes the optimal sampling strategies for diverse data drift types. 
Specifically, \sysname calculates the time interval $\Delta t$ from the data drift start to the end.
And then it computes the distance $d$ in data distribution between the first half of the data during the data transformation (\ie $[{t}_{1}, \frac{{t}_{1} + {t}_{2}}{2}]$) and the historical data (\ie $win_1$) to identify the type of data drift at the moment $t$ $\pi (t)$ as sudden, incremental and gradual as follows:
\begin{equation}
\pi (t)= \begin{cases}\mathrm{sudden} 
  & \Delta t < \tau \\\mathrm{incremental} 
  & \Delta t \ge  \tau \ and \ d>{d}_{0} \\\mathrm{gradual} 
  & \Delta t \ge  \tau \ and \ d\le {d}_{0}
\end{cases}
\end{equation}
When $\Delta t = {t}_{2} - {t}_{1}$ (detected in \S\ref{sec:422}) is short enough (\eg $\tau =$ 90~s), the current drift type is considered as sudden drift. 
Otherwise, it is the incremental or gradual drift.
To further differentiate, 
we introduce the term "intermediate concept"~\cite{bib:csur2014:survey}, which refers to the distribution of the data during data drift (\ie, the interval $[{t}_{1},{t}_{2}]$).
As shown in Figure~\ref{fig:drift_type}, when $d$ is more than threshold ${d}_{0}$, the data distribution within the first half of data transformation is approximate to the historical one, and the intermediate concept is one of the old or new data (\eg incremental drift). 
On the contrary, if $d$ is below the threshold, the intermediate concept is the old or new data (\eg gradual drift).
10We note that we leverage the frame difference to measure the distance between the data distributions~\cite{bib:sensys2015:chen}. 
And we set the threshold ${d}_{0}$ as 0.2 times the distance between the old and new data distributions.

Based on the identified data drift types, we employ the specialized video frame sampling strategy for them.
Since the purpose of model evolution is to better deal with the current or upcoming new scenes, it is unnecessary to pay  much attention to the historical data during retraining, which also leads to longer retraining delay and lag accuracy gain~\cite{bib:iccv2019:Mullapudi}. 
As discussed in \S\ref{sec:42}, we trade off the number of video frames uploaded and the DNN retraining time by selecting the suitable video frames from the interval $[{t}_{1},{t}_{3}]$.
The sampling strategies are as follows:

\textbf{Sudden drift}.  
We use a fixed sampling rate of ${r}_{f}$ (\eg 0.6~fps) to sample the video frames, because of the sudden occurrence of drift, moving from the old data distribution to a new one, with a more uniform distribution of data within the interval $[{t}_{1},{t}_{3}]$.
In addition, the fixed sampling rate will reduce the time of frame selection, which leads to less extra time overhead in evolution.

\textbf{Incremental drift}. 
Since the data changes continuously during the data conversion process, \sysname uses a linear sampling rate ${r}_{t}$ as below.
\begin{equation}
r_{t}=  \mathrm{MIN} (r_{max},r_{0}+floor (\frac{t-t_{1}}{30} )\cdot \Delta r)
\end{equation} 
where $r_{max}$ indicates the maximum value of the sampling rate, ${r}_{0}$ denotes the initial value of the sampling rate, the $floor()$ present the downward rounding function (\ie $floor(x)=N$, if $N\le x< N+1$ and $N\in \mathbb{Z}$), $\Delta r$ is the increment of the sampling rate.
We set ${r}_{max}$, $r_{0}$ and $\Delta r$ as 1~fps, 0.1~fps and 0.05~fps by default, respectively.

\begin{figure*}[t]
    \centering
    \includegraphics[width=.85\textwidth]{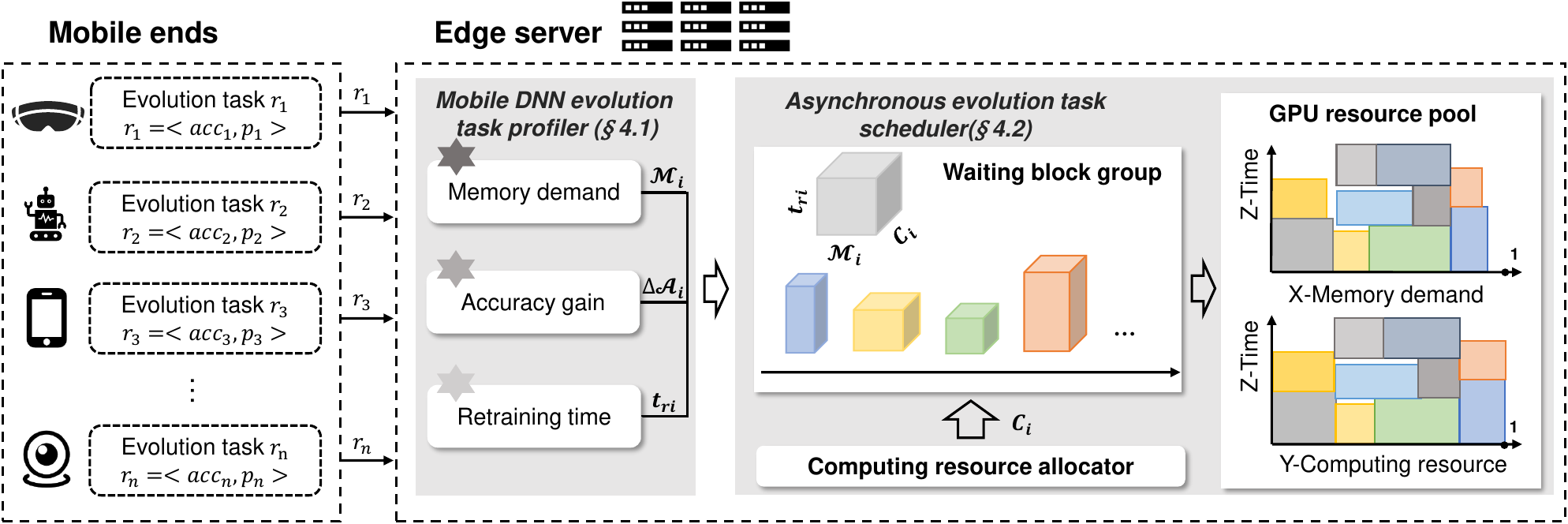}
    \caption{Illustration of asynchronous DNN evolution task scheduling.}
    \vspace{-3mm}
    \label{fig:design_2}
\end{figure*}

\textbf{Gradual drift}. 
\sysname first uses a fast and straightforward frame difference method, \ie the absolute value of the pixel difference~\cite{bib:sensys2015:chen}, to remove redundant frames whose difference is lower than the threshold $1280*720*\epsilon_1$.
Then it performs a feature comparison to judge the difference between the non-redundant frame and the global view of the data distribution to predict the contribution of every frame to the retraining and only uploads the most representative frames whose contribution exceeds a threshold $\epsilon_2$.
To extract the knowledge from the global view of existing data distribution, which is non-observable by the local mobile end, the server learns a conditional distribution $\mathcal{Y} \to \mathcal{Z}$ via a 2-layer conditional generator $G$~\cite{bib:icml2021:zhu} and broadcasts it to mobile ends periodically.
Here $\mathcal{Z}$ is the latent feature space, and $\mathcal{Y}$ is the output space.
For example, the mobile end generates features $\boldsymbol{z}$ and inference results $y$ for each frame. 
With this generator $G$, the mobile end can get the feature distribution $\boldsymbol{z'}=G(y)$ in the global view. 
When the difference $\mathcal{D}(\boldsymbol{z},\boldsymbol{z'})$ between $\boldsymbol{z}$ and $\boldsymbol{z'}$ is greater than $\epsilon_2$, we consider it a representative frame with a significant deviation.
The difference $\mathcal{D}(\boldsymbol{z},\boldsymbol{z'})$ is defined as:
\begin{equation}
\mathcal{D}(\boldsymbol{z},\boldsymbol{z'}) = \frac{1}{C}\sum_{c=1}^{C}\mathcal{D}^c(\boldsymbol{{z}^c},\boldsymbol{z^{c'}}) 
\end{equation}

\begin{equation}
\mathcal{D}^c(\boldsymbol{{z}^c},\boldsymbol{z^{c'}}) = \frac{1}{K}\sum_{k=1}^{K}\mathcal{D}^c_k(\boldsymbol{{z}^c},\boldsymbol{z^{c'}})
\end{equation}

\begin{equation}
\mathcal{D}^c_k(\boldsymbol{{z}^c},\boldsymbol{z^{c'}}) = \sqrt{\sum_{i=1}^{n}\left ( z^{c}_{ik} - z^{c'}_{ik} \right )^2  }
\end{equation} 
Where $\mathcal{D}^c_k(\boldsymbol{{z}^c},\boldsymbol{z^{c'}})$ denotes the feature difference of the $k$-th bounding box of category $c$ (we use the Euclidean distance of the vector to measure the difference between features), and $\mathcal{D}^c(\boldsymbol{{z}^c},\boldsymbol{z^{c'}})$ denotes the average difference of category $c$ in that video frame.

\section{\sysnameposs Edge-server Design}
\label{sec:edge}

Multiple mobile DNN evolution tasks asynchronously arrive at the edge server, as shown in Figure \ref{fig:design_2}.
The \textit{primary bottleneck} to extend the evolving process from the aforementioned single-mobile DNN case to the multi-mobile DNNs case is the edge server's limited resources (\eg GPU memory and computing resource).
It restricts the optimization of diverse tasks' scheduling $t_s$ and retraining time $t_r$.
This section presents the task scheduling mechanisms to maximize the average QoE for multiple mobile users.
Specifically, the edge server utilizes the \textit{mobile DNN evolution task profiler} (\S \ref{s_estimator}) to estimate evolution tasks' performance metrics.
These estimated metrics then are input to the \textit{asynchronous task scheduler} (\S \ref{s_task_scheduler}) to schedule multiple tasks adaptively.
In addition, the edge server adopts the \textit{sparse DNN retraining} (\S \ref{sec:retrain}) mechanisms to reduce the unimportant parameters of evolved DNN and the retraining data for shortening the retraining time of each mobile DNN.

\subsection{Mobile DNN Evolution Task Profiler}
\label{s_estimator}

For the asynchronous evolution task scheduler to perform suitably, it is essential to provide an accurate and timely estimate of each evolution task's performance metrics before conducting the retraining.
The metrics include the evolution task's GPU memory demand $\mathcal{M}$, the retrained model's accuracy gain $\Delta \mathcal{A}$, and the retraining time $t_r$.

\parahead{Memory Demand $\mathcal{M}$}
Each evolution task's GPU memory demands $\mathcal{M}$ consist of the model parameters $m_p$, the intermediate results $m_f$ (\ie feature map), the back-propagation gradients $m_g$, the optimizer's parameters $m_{opt}$ (\eg SGD optimizer~\cite{bib:sgd:Robbins}), and the in-layer storage of cuDNN workspace $m_{ws}$ used to support the layer calculation~\cite{bib:isca2018:jain}, \ie
$\mathcal{M}_i = m_{p_i} + m_{f_i} + m_{g_i} +m_{{opt}_i} + m_{{ws}}$.
Specifically, the model parameter size $m_p$ is the sum of all layers (\eg convolutional, fully-connected, and BN layer), \ie the parameter number multiplied by the bit widths.
The model parameter number can be directly derived from the layer's architecture, \eg ${{m}_{s_{conv}}}=({C}_{in}\times{C}_{out}\times{K}_{1}\times{K}_{2})\times \mathcal{B}_p$.
The bit-width $\mathcal{B}_p$ is determined by the parameter quantization settings on a specific mobile end, \eg 8-bit, 16-bit, or 32-bit.
%
%
And the memory occupation of intermediate features can also be directly calculated, given the input size and the layer's architecture. 
For example, 
${m}_{f_{conv}}={\frac{{w}_{in}-{K}_{1}+{2}\times{p}_{1}}{{s}_{1}+{1}}}\times{\frac{{h}_{in}-{K}_{2}+{2}\times{p}_{2}}{{s}_{2}+{1}}}\times{C}_{out}\times \mathcal{B}_{p}$.
Since the gradients are derived from intermediate features, their memory demands are approximatively equal to that of the intermediate features, \ie $m_{g}= m_f$. 
The optimizer's memory demand $m_{opt}$ is proportional to the model parameter size because the optimizer is used to update all parameters.
We empirically set the ratio as $2 \sigma_2$, \ie $m_{opt} = 2 m_{p}$.  
The memory demand of workspace $m_{ws}$ can be profiled offline for a given server platform, \eg 
we empirically set $s_{ws} = 847.30$MB for the NVIDIA server with CUDA11.1 and cuDNN8.0.5.

\parahead{Accuracy Gain $\Delta \mathcal{A}$}
It is challenging to estimate the accuracy gain of retraining in advance because it is relative to diverse DNNs and heterogeneous training data. Both of them are dynamic.
Collecting massive retraining records for offline regression fitting is costly and prone to inaccuracy.  
To this end, we propose to fit a curve \textit{online} with the training progress to account for the dynamics of deep models and retraining data.
%
We referred to the fitting method in~\cite{bib:EuroSys2018:peng}, which has been verified in the training assessment.
To reduce the overhead of online profiling, we adopt a small number of video frames (\eg 10$\%$) and a few training epochs (\eg five epochs) to fit a nonlinear curve of training epochs and inference accuracy. 
Specifically, we test the model accuracy after each epoch using the mini-batch data and collect the paired accuracy and epoch number. 
A non-negative least squares (NNLS) solver\cite{scipy} is used to compute the coefficients involved in the curve.
The online fitting process is fast and real-time, \eg a few milliseconds, and the occupied memory is quickly released.
Then, the edge server uses this curve to predict each retraining model's accuracy using the estimated epoch number of retraining convergence and obtain its accuracy gain $\Delta \mathcal{A}$.

\parahead{Retraining Time $t_r$}
We leverage a parametric regression module with three fully-connected layers for building the map between the retraining time and model retraining factors.
The primary factors that feature a model's retraining time include the model size, the amount of retraining data, the number of model layers participating in the retraining (\ie unfrozen layers), the retraining epochs, and the batch size. 
And we collect $200$ samples of diverse factor settings and the corresponding retraining time to train the regression network offline. 
As we will show in \S \ref{subsub_profiler_time}, this regression network is compact and efficient and generates the estimation result quickly, \eg $\leq 0.09 ms$.

\begin{figure}[t]
    \centering
    \includegraphics[width=.36\textwidth]{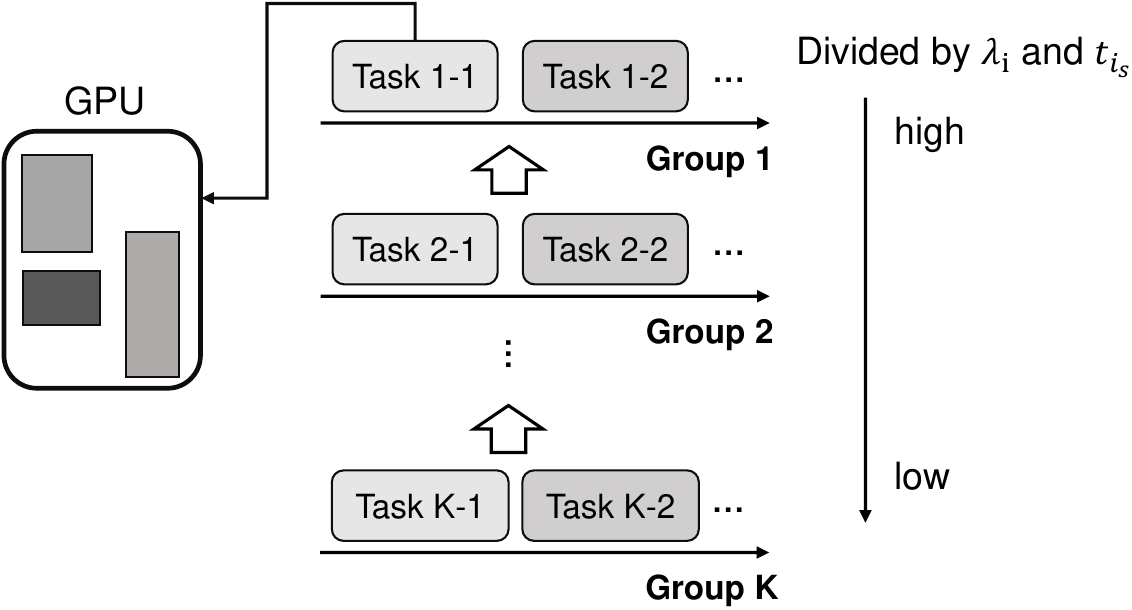}
    \caption{Illustration of evolution task grouping mechanism.}
    \label{fig:design2}
\end{figure}

\subsection{Asynchronous Evolution Task Scheduler}
\label{s_task_scheduler}
It is intractable to schedule the asynchronous mobile DNN evolution (\ie retraining) tasks for minimizing the $T_{retrain}$ in Equation~\ref{eqn:t_retrain} and maximizing the overall QoE $Q_{avg}$ in Equation~\ref{equ:opt}, which is a two-way dynamic inter-allocation problem of resource supply and demand.
The reasons are two-fold:
\textit{(i)} An evolution task can only be executed if it obtains enough memory resources, \ie more than its memory demand $ \mathcal{M}_i$. 
We can only control different tasks' scheduling time $t_{s}$ by determining when to allocate enough memory resources for them;
\textit{(ii)} The more computing resources an evolution task obtains, the shorter retraining time it will have.
We can tune each task's retraining time $t_r$ by adjusting the amount of allocated computing resources $\mathcal{C}_i$.
Based on the above observations, 
our principle for global task scheduling is to \textit{maximize the Quality of Experience per unit of edge resource}.
That is, the edge server preferentially schedules the evolution tasks with larger model evolving urgency $\lambda_i$ mentioned in \S\ref{sec:form} which is a function of the current inference accuracy $\mathcal{A}_i(t)$ and accuracy drop $\Delta \mathcal{A}$, meanwhile minimizing the scheduling time $t_{i_s}$ and retraining time $t_{i_r}$. 
%
Due to different mobile DNN evolving urgency $\lambda_i$ and limited edge resources, \sysname should tackle the trade-off between reducing the ${t_{i_s} + t_{i_r}}$ of the specific tasks and all tasks.
Moreover, it is NP-hard to quickly select the appropriate task combination from all asynchronous requests with different resource demands to fit them into the GPU with dynamic resource availability and maximize the average QoE. The reason is that this problem can be reduced to the Knapsack problem and the solution to it can be verified in polynomial time.

To this end, we transform the evolution task scheduling problem as a "Tetris stacking problem", shown in Figure~\ref{fig:design_2}.
Specifically, when putting a task into the GPU, its retraining time $t_{i_r}$, memory demand $\mathcal{M}_i$, and computing resources $\mathcal{C}_i$ can be described as a cuboid block. 
Therefore, the task scheduler needs to select the most valuable combination of blocks from a suitable block group to cover the area of the GPU resource pool as much as possible without collision among task blocks.
Thus, we develop an evolution task grouping mechanism (\S \ref{sec:searchspace}) to obtain a well-designed block group for selecting task blocks, thereby overcoming differences in model evolving urgency $\lambda_i$ among tasks and improving the search efficiency for task selection.
In addition, we employ a dynamic programming-based algorithm (\S \ref{sec:taskselection}) to select the optimal evolution task combination and utilize the adaptive edge resource allocator to allocate suitable resources (\S \ref{s_resource_allocator}) to the task blocks and thereby minimize the average $t_{i_s}+t_{i_r}$ and improve the overall QoE $Q_{avg}$ to boost the search efficiency.

\subsubsection{Well-designed Search Space}
\label{sec:searchspace}
A well-designed search space for selecting task combinations is important to tackle the attribute difference among multiple evolution tasks caused by the model evolving urgency $\lambda_i$, and improve the search speed of scheduling. 
Because a suitable search space can avoid considering the differences in $\lambda_i$ among tasks and reduce the number of searches, thereby increasing its speed.
We optimize the search space of task selection by an evolution task grouping mechanism, as shown in Figure~\ref{fig:design2}.
Specifically, we group all tasks into $K$ groups according to model evolving urgency $\lambda_i$ from high to low to make tasks in the same group have similar urgency without considering scheduling time $t_{i_s}$.
Thus, optimizing $Q_{avg}$ is transformed to maximizing $Q_{avg} = \frac{1}{K} \sum_{m=1}^K Q_{avg}^m $, where $Q_{avg}^m$ represents the average QoE of the $m$-th group. 
And $Q_{avg}^m$ is refined as:
\begin{equation}
\begin{aligned}
    Q_{avg}^m = \frac{1}{n_{m}} \sum_{i=1}^{n_{m}} \lambda_i Q_i
= \frac{1}{n_{m}} \times \lambda_{avg_m} \times \sum_{i=1}^{n_{m}} Q_i
\label{equ_Qopt2}
\end{aligned}
\end{equation}

Where ${n_{m}}$ denotes the number of tasks in the $m$-th group and $\lambda_{avg_{m}}$ presents the average value of the task's model evolving urgency in the $m$-th group. 
In summary, we decompose the optimization problem of $Q_{avg}$ into several optimization problems of $Q_{avg}^m$, and then we can obtain the optimal task combination and maximize $Q_{avg}^m$ and overcome the differences in model evolving urgency among tasks.

\begin{figure}[t]
    \centering
    \includegraphics[width=.36\textwidth]{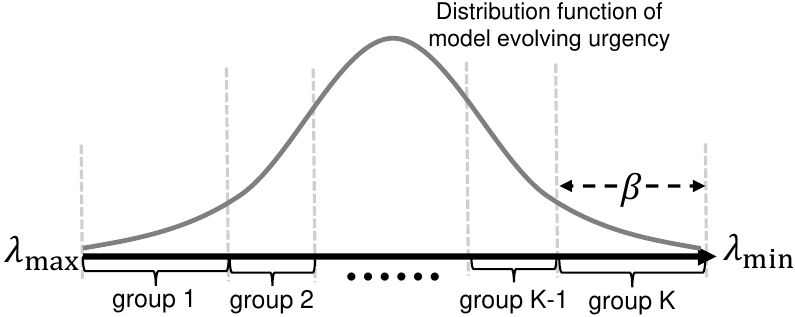}
    \caption{Adaptive group number decision.}
    \label{fig:number}
\end{figure}

Conducting task grouping needs to consider a finer-grained aspect, \ie the task grouping probability, which is related to each group's task number and evolving urgency range. In particular, we have two choices for task grouping strategy: equal task probability grouping or equal urgency range grouping. 
Equal task probability means that tasks have an equal probability of appearing in each group, while equal urgency range grouping represents the urgency range of each group as the same.
As we will show in \S~\ref{subsec:exp_555}, we adopt the equal task probability grouping by default because it can achieve better performance.

Moreover, the appropriate setting of group number $K$ is highly significant.
The more groups are divided, the smaller the search space for task selection. 
And the fewer group numbers may diminish the effect of search speedup.
We find that the appropriate group number varies with the hardware level of the edge server. 
When the edge server's hardware resources are abundant, it can serve plenty of mobile ends, thereby dealing with more evolution requests. 
And the group number can be appropriately increased. 
Otherwise, the group number should be reduced.
Therefore, we design the adaptive group number setting strategy according to the edge server hardware level, as shown in Figure~\ref{fig:number}.
Given an edge server that can deal with $N$ requests concurrently at most, the group number $K$ can be calculated as below.

\begin{equation}
    K = int(\frac{2}{1-erf(\frac{\frac{1}{2}(\lambda_{max}-\lambda_{min})-\beta}{\sqrt{2}\sigma})})
\label{equ:m}
\end{equation}

\begin{equation}
    \frac{1}{2}N[1-erf(\frac{\frac{1}{2}(\lambda_{max}-\lambda_{min})-\beta}{\sqrt{2}\sigma})] \geq n_{min}
\label{cons1}
\end{equation}

\begin{equation}
    \beta \leq \varepsilon
\label{cons2}
\end{equation}
Where $int()$ denotes the rounding function, $erf()$ presents the integral of a normal probability density function, $\beta$ indicates the model evolving urgency range length of the last group, $n_{min}$ is the minimum number of tasks in each group, and $\varepsilon$ is the threshold of acceptable model evolving urgency range length.
Since the model evolving urgency from the mobile side is related to the mobile application demands, which may change randomly, we approximate the value distribution of the model evolving urgency to a normal distribution(according to the central limit theorem~\cite{bib:central_limit}) with mean $\frac{1}{2}(\lambda_{max}+\lambda_{min})$ and variance $\sigma^{2}$. 
Here, $\lambda_{max}$ and $\lambda_{min}$ denote the maximum value and the minimum value of model evolving urgency, respectively.
Based on the normally distributed data characteristics, we calculate group number $K$ in Equation~\ref{equ:m} under the constraints of Equation~\ref{cons1} and Equation~\ref{cons2}.
Equation~\ref{cons1} limits the minimum number of tasks per group, and Equation~\ref{cons2} presents that the urgency range of the last group and the first one, which is the largest among all groups, is within our tolerance interval to limit the maximum number of tasks in each group. Based on the grouping strategy and adaptive group number, we can obtain the model evolving urgency ranges for each group.
When new tasks arrive at the edge server asynchronously, we still group them by model evolving urgency according to the original urgency range to prevent the starvation of ones with low model evolving urgency.


\subsubsection{Mobile DNN Evolution Task Selection}
\label{sec:taskselection}

\begin{algorithm}[t]
  \caption{Mobile DNN Evolution Task Selection} 
  \label{alg:task_sele} 
  \scriptsize
  \KwIn{evolution tasks $(r_{1},r_{2},...,r_{N})$,
  the predicted completed time of tasks in GPU
  $(t_1,t_2,...,t_i,...,t_m)$,
  the retraining time of tasks in GPU
  $(t_{1_r},t_{2_r},...,t_{i_r},...,t_{m_r})$
  the time value of each task $(v_{1},v_{2},...,v_{N})$, the memory demand of each task $(\mathcal{M}_{1},\mathcal{M}_{2},...,\mathcal{M}_{N})$ and GPU memory capacity $\mathcal{M}_{c}$.}  
  \KwOut{the $n-$ary vector $(\varphi_{1},\varphi_{2},...,\varphi_{N})$.}  
  
  Initialize \thinspace $\Delta t \gets \beta \cdot t_{i_r}$
  
  \If{$t_i + \Delta t \geq t_{i+n}$}{$\mathcal{M}_{c} = \mathcal{M}(t_{i+n})$} 
  \Else{
  $\mathcal{M}_{c} = \mathcal{M}(t_{i})$
  }

  Create \thinspace $dp[0:N][0:\mathcal{M}_{c}+1]$

  Initialize \thinspace $dp$ \thinspace with \thinspace $\mathcal{M}_{1}, v_{1}$

  \For{$k = 1 \to N-1$}{
    \For{$rm = 1 \to \mathcal{M}_{c}$}{
      \If{$\mathcal{M}_{k} > rm$}{$dp[k][rm] \gets dp[k-1][rm]$}
      \Else{$dp[k][rm] \gets \textbf{MAX}(dp[k-1][rm],dp[k-1][rm-\mathcal{M}_{k}]+v_{k})$}}}

  Initialize \thinspace $(\varphi_{1},\varphi_{2},...,\varphi_{N}) \gets 0$

  \While{$N > 1$}{
    \If{$dp[N-1][\mathcal{M}_{c}] != dp[N-2][\mathcal{M}_{c}]$}{$\varphi_{N-1} \gets 1,\mathcal{M}_{c} \gets \mathcal{M}_{c} - \mathcal{M}_{N-1}$}
    $N \gets N-1$}
  \If{$dp[0][\mathcal{M}_{c}] > 0]$}{$\varphi_{0} \gets 1$}
  \textbf{return} the $n-$ary vector $(\varphi_{1},\varphi_{2},...,\varphi_{N})$
\end{algorithm}  

Based on the above search space, we propose a dynamic programming-based task selection algorithm to pick up the optimal task combinations.
Essentially, the maximum number $N$ of DNN evolution (\ie retraining) tasks that the edge server can support concurrently.
This is dynamically determined by each task's memory demands and the availability of memory supply $\mathcal{M}_s$, \ie $\sum_{i=1}^{N}{\mathcal{M}_i} \leq \mathcal{M}_s$.
Therefore, the \sysnameposs edge server \textit{selects the optimal task subset from to-be-scheduled evolution tasks in the most urgent group for the best overall performance benefits.} 
To achieve this goal, \textit{(i)} \sysname makes full use of the GPU's available memory and computing resources can maximize the throughput of the GPU, and reduce the retraining time of tasks being served and the scheduling time of subsequent ones;
\textit{(ii)} \sysname utilizes the short job first principle can effectively shorten the average scheduling time of all tasks, thereby further reducing $t_r+t_s$. 
Therefore, we define time value $\frac{\alpha}{t_{i_r}}$ as the item value in the context of the evolution task selection problem, which means shorter tasks obtain greater values. 
Then, the problem turns into finding the n-ary vector $(\phi_1, \phi_2, ..., \phi_i, ..., \phi_n), \phi_i \in 	\left\{0,1\right\}$, for maximizing the total value $\sum_{i=1}^{N}\phi_i\frac{\alpha}{t_{i_r}}$ and satisfying the constraint conditions $\sum_{i=1}^{N}\phi_i{\mathcal{M}_i} \leq \mathcal{M}_s$ and $\sum_{i=1}^{N}\phi_i{\mathcal{C}_i} \leq \mathcal{C}_s$.

The dynamic programming-based evolution task selection algorithm is outlined in Algorithm \ref{alg:task_sele}.
First, to determine the maximum capacity of the GPU resource pool, which is formulated as the edge server's current available memory capability $\mathcal{M}_c$, we employ the prediction-based memory decision approach. 
Specifically, at time $t_i$ when the $i$-th task is completed, we need to judge whether there will be other tasks completed within $\Delta t$, which is formalized  as $t_i + \Delta t \geq t_{i+n}$ where $\Delta t = \beta \cdot t_{i_r}$ and $t_{i+n}$ is the completion time point of the $(i+n)$-th task.
If no other tasks are completed during this period, the currently available memory $\mathcal{M}_c = \mathcal{M}(t_i)$. Where $\mathcal{M}(t_i)$ is the remaining available memory at $t_i$, the selected tasks are put in GPU now. Otherwise, the memory $\mathcal{M}_c = \mathcal{M}(t_{i+n})$ and correspondingly, the time for tasks to be scheduled and served is postponed to $t_{i+n}$. 
Using the above memory decision method, the edge server can avoid frequent task scheduling and prevent the starvation of tasks with large memory demands, thereby meeting the needs of mobile ends and ensuring the system's normal operation.

Then, we define a two-dimensional array $dp$, each item $dp[k][rm]$ represents the maximum sum of the time value selected from $k$ evolution tasks under the constraint of remaining memory $rm$.
For each $dp[k][rm]$, if the memory demands of the $k-$th evolution task are higher than the currently available memory supply $rm$, the $k-$th evolution task cannot be selected, \ie $dp[k][rm]=dp[k-1][rm]$.
Otherwise, if the memory demands of the $k-$th evolution task are lower than or equal to the available memory resource supply $rm$, we further compare the sum of values obtained by selecting and not selecting the $k-$th task, \ie $dp[k-1][rm-\mathcal{M}_k] + v_k$ and $dp[k][rm]$, assign the larger value to update $dp[k][rm]$.
The algorithm iteratively updates the $dp$ array until it obtains $dp[N][\mathcal{M}_c]$, denoting the maximum sum of time value selected from $N$ evolution tasks in group 1 with the satisfaction with the edge server's current available memory capability $\mathcal{M}_c$. Finally, we can obtain the n-ary vector $(\phi_1, \phi_2, ..., \phi_i, ..., \phi_n)$, which presents the selected task combination.
After all tasks in group 1 are completed, the tasks in the $i$-th (i = 2, 3,..., K) group move up to the $(i-1)$-th group, and we continue employing the dynamic programming-based evolution task selection algorithm to select tasks from the current group 1.

\subsubsection{Adaptive Edge Resource Allocator}
\label{s_resource_allocator}
The edge server employs two mechanisms to allocate suitable computing and memory resources to the selected tasks.

\parahead{On-demand Memory Resource Allocation}
The edge server allocates memory resources for each selected task \textit{on-demand} based on the memory demands estimated by the mobile DNN evolution task profiler. 
There are two reasons: \textit{(i)} The memory resource is the bound for an evolution task, \ie only when the allocated memory resources are higher than the evolution task's memory demand can the GPU execute it. 
\textit{(ii)} Extra memory resources exceeding the task's demand $ \mathcal{M}_i$ cannot bring any additional performance benefits.

\parahead{Demand-driven Computing Resource Allocation}
By default, when multiple mobile-end model evolution tasks compete for limited computing resources, the edge server allocates the same computing resources to each parallel process, which will not maximize their utilization.
 Based on the previous memory usage analysis for model retraining, we find that tasks with a larger number of model parameters and more intermediate results take up more memory resources. Usually, such tasks will be accompanied by more calculations, and their utilization rates of computing resources per unit of time are also higher. In addition, these evolution tasks with large memory demands will block the opportunity to reallocate resources for other waiting tasks. Therefore, to improve resource utilization and avoid blocking problems, \sysname allocates the computing resources for each evolution task based on their memory demands $ \mathcal{M}_i$. 
The principle is to allocate more computing resources for heavy evolution tasks with larger memory demands. 
Specifically, we model the amount of allocated computing resource $\mathcal{C}_i$ for the $i-$th evolution task as:
$\mathcal{C}_{i}=\frac{\mathcal{M}_i}{\sum_{i=1}^{N}\mathcal{M}_i} \times \mathcal{C}_s(t)$.
Where $N$ is the number of evolution tasks being executed on the edge server, and $\mathcal{C}_s(t)$ denotes the dynamic availability of GPU computing resources.

Technically, we adopt the NVIDIA Volta Multi-Process Service (MPS) API~\cite{voltamps} to allocate the memory and computing resources adaptively. 
It transparently enables cooperative multiple CUDA processes for concurrently executing numerous evolution tasks on the NVIDIA GPU.

\subsection {Sparse DNN Retraining}
\label{sec:retrain}
\subsubsection{Bounding Box-level Sample Filtering}\label{s:label_gen}
DNN retraining for evolution requires labeled data, while the video frames uploaded by the mobile ends are unlabeled.
To generate labels, \sysname uses a golden model on the edge server to perform inference on each frame 
and uses the obtained class probabilities and bounding boxes as the pseudo-label.
To reduce the retraining time $t_{r}$ in Equation \ref{eqn:t_retrain}, \sysname filters the uploaded video data in a fine-grained manner.
Specifically, the class probability is a good indicator for the bounding box accuracy~\cite{bib:arXiv2021:lu}.
\sysname utilizes a \textit{bounding box-level} pseudo label filter to filter out those bounding boxes with low confidence and only use the labeled "bounding boxes" with high confidence for retraining, as shown in Figure~\ref{fig:design_1}.
The contents in these low-confidence bounding boxes are set as the background, 
preventing them from participating in the retraining process.
We set the filing threshold as $0.5$ by default, and experimentally validate the efficiency of the proposed filter in \S \ref{subsec:exp_642}.

\begin{figure}[tp]
    \centering
    \includegraphics[width=.4\textwidth]{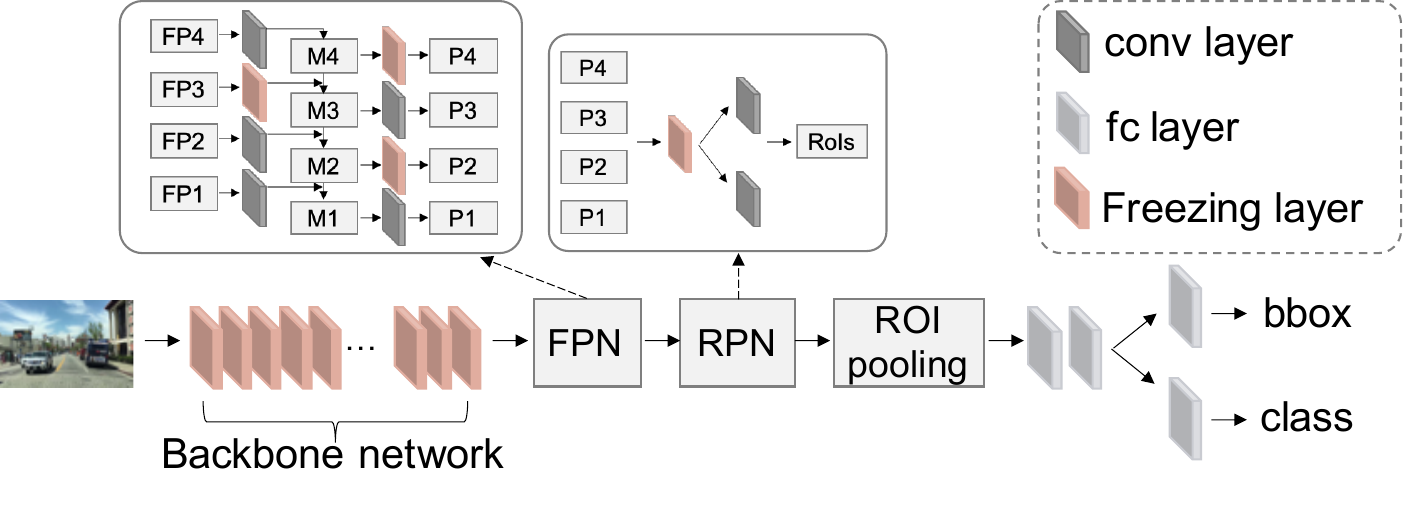}
    \caption{Showcase of freezing retraining Faster R-CNN.}
    \label{fig:faster-rcnn}
\end{figure}

\subsubsection{Compression-aware DNN Freezing Retraining}
\label{subsec:train_compress}
To balance the DNN retraining efficiency, \ie reducing the retraining delay and improving effect, \ie improving the accuracy of the retrained model, \sysname adopts the compression-aware DNN freezing retraining strategy.
Specifically, two observations motivate our design.
First, the layers inside a neural network gradually transition from task-independent to task-specific from the first to the last ~\cite{bib:yosinski2014:transferable}. 
For example, an object detection model like Faster R-CNN contains a backbone network, 
feature pyramid network (FPN), region proposal network (RPN), RoI pooling, and classification layer, as shown in Figure~\ref{fig:faster-rcnn}.
Freezing the task-independent layers during retraining introduces minimum harmful effect on accuracy~\cite{bib:eccv2020:isikdogan},
but significantly reduces the computational overhead and saves training time. Therefore, we freeze the backbone when retraining networks like Faster R-CNN.
Second, a layer's information includes task-specific information and redundant information~\cite{bib:KDD21:gao}.
One of the major tasks for DNN compression is to discover layers that contain redundant information. %
Therefore, we apply similar techniques to find the redundant layers and freeze them during the retraining. 
Specifically, we pre-generate the optimal quantization strategy for diverse mobile ends offline. And we approximate the amount of each layer's non-redundant information 
and progressively freeze those redundant layers in FPN and RPN modules.

\section{Evaluation}
\label{sec:experiment}
This section presents the experimental settings and the comprehensive system performance of \sysname. 

\begin{table}[]
\centering
\scriptsize
\caption{Summary of testing mobile data.}
\begin{tabular}{|c|c|c|c|}
\hline
\textbf{Mobile collected videos} & \textbf{City}        & \textbf{Time quantum} & \textbf{Duration} \\ \hline
\textbf{D1}       & A     & Dusk        & $30$~min    \\ \hline
\textbf{D2}       & A     & Night       & $30$~min    \\ \hline
\textbf{D3}       & B & Daytime     & $30$~min    \\ \hline
\textbf{D4}       & C & Daytime     & $30$~min    \\ \hline
\end{tabular}
\label{tb:videodata}
\end{table}

\subsection{Experimental Setups}
\label{sec_exp_setup}

\parahead{Implementation}
We implement \sysname with PyTorch~\cite{bib:nips2019:Paszke} and MMDetection~\cite{bib:arXiv2019:Chen} in Python.
We use ten development boards (\eg Raspberry Pi3, Raspberry Pi4) equipped with mobile robots as the mobile ends and two NVIDIA Tesla V100 GPUs with 16GB memory and two NVIDIA GeForce RTX3080 GPUs with 10GB memory as the edge servers. 
By default, multiple evolution tasks are executed directly on the NVIDIA GPU, using time-slice rotation. 
\sysname uses Volta MPS~\cite{voltamps} to adaptively allocate GPU resources for multiple evolution tasks, realizing parallel execution.

\parahead{Datasets and model configurations}
We experiment with four videos collected by real-world mobile vehicles at the diverse time (\ie dusk, night, and daytime) in three cities for testing. 
Each testing data experiences diverse mobile scenarios and undergoes different types of data drift. 
Table \ref{tb:videodata} lists the details of these datasets.
And the public COCO~\cite{bib:Springer2014:Lin} and BDD~\cite{bib:cvpr2020:bdd} datasets are used for model pre-training.

We employ four object detection DNNs, \ie Faster RCNN with ResNet50~\cite{bib:arxiv1016:resnet} (model 1) and  MobileNetV2~\cite{bib:cvpr2018:mobilenetv2} backbone network (model 2), YOLOv3~\cite{bib:arXiv2018:Redmon} with Darknet53~\cite{bib:arXiv2018:Redmon} (model 3) and ResNet50~\cite{bib:arxiv1016:resnet} (model 4) backbone network.
Besides, we use two types of compression techniques (\ie quantization and pruning) with diverse compression ratios to compress the above DNNs.

\parahead{Comparison baselines} 
We compare \sysname with the following baselines in the single mobile end case:

 \begin{itemize}
    \item \textbf{Original compressed model (A1).} The mobile end loads the object detection model pre-trained by the public dataset COCO~\cite{bib:Springer2014:Lin}. 
    \item \textbf{Domain adaptation (A2)~\cite{bib:cvpr2018:domain}.} The edge server retrains DNNs to adapt to new data using the adversarial training-based domain adaptation method.
    \item \textbf{Down-sampling method (A3)~} The mobile end uploads video frames with a sample rate of $0.06$ to the edge server for model evolution.
    \item \textbf{Cloud-assisted model evolution (A4)~\cite{bib:iccv2021:Khani}.} The mobile end continuously transmits video data to the cloud server for model evolution.
\end{itemize}

And we leverage the following baselines for comparisons in the case of multiple asynchronous mobile ends.

\begin{itemize}
    \item \textbf{Default GPU scheduling (B1).} The edge server executes evolution tasks by default.
    \item \textbf{Serial execution w/o priority (B2).} The edge server 
    schedules multiple evolution tasks in sequence according to their arrival order. 
    \item \textbf{Serial execution w/ priority (B3).} The edge server 
    schedules multiple evolution tasks in sequence according to their model evolving urgency. 
    \item \textbf{\sysname w/o priority (B4).} \sysname's task scheduling method without task grouping mechanism.
\end{itemize}

\begin{figure}[t]
    \centering
    \includegraphics[width=.44\textwidth]{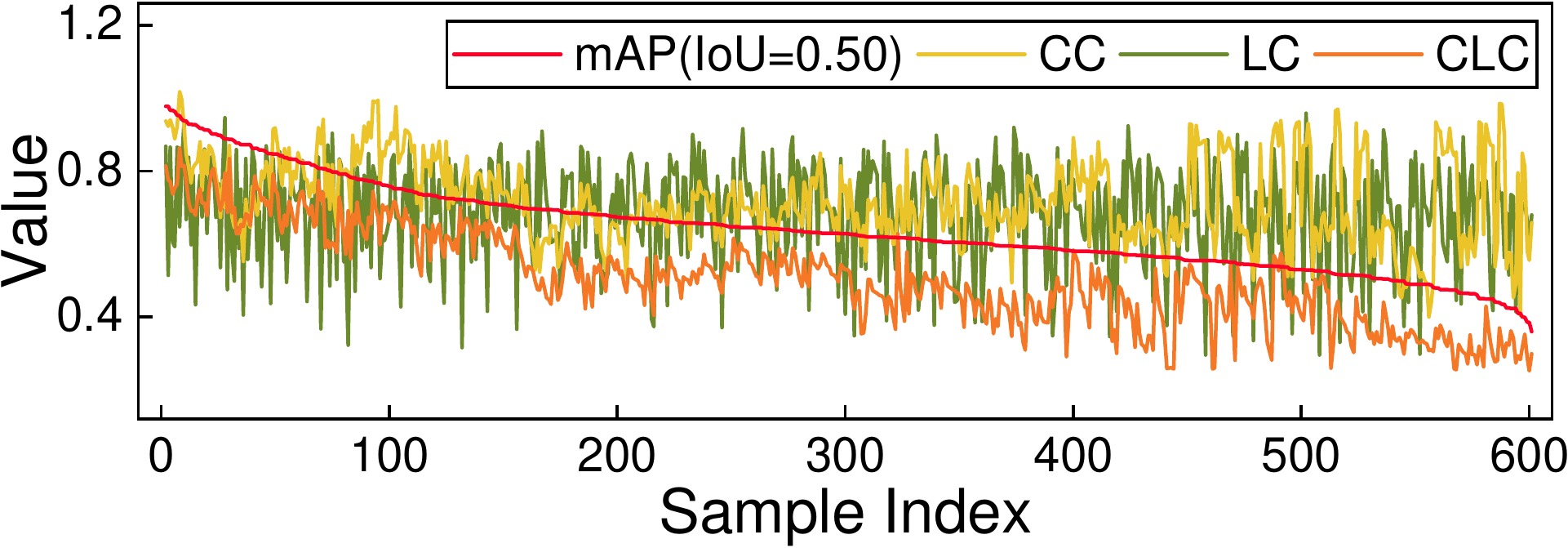}
    \caption{Correlation between the accuracy of the model and three different metrics}
    \label{fig:exp:211}
\end{figure}

\subsection{Limitations of Existing Methods}
\label{s_motivation}
We first experimentally elucidate the key limitations of existing methods, validating the demands on \sysnameposs design.

\subsubsection{Different Metrics for Measuring Accuracy Drop at the Mobile End}\label{exp:211}

To demonstrate that the metric chosen by \sysname accurately reflects the accuracy drop of the model, 
we select $600$ samples and calculate the accuracy(mAP(IoU=0.50)), classification confidence($CC$), localization confidence($LC$), and detection confidence($CLC$) used by \sysname for each sample data separately.
Figure~\ref{fig:exp:211} shows the correlation between the accuracy(mAP(IoU=0.50)) and three different metrics.
We can see that there is a stronger positive correlation between the accuracy(mAP(IoU=0.50)) and detection confidence, 
and the other two metrics do not reflect the detection accuracy very well. 
According to the experimental results, we have three findings.
Firstly, the classification confidence is still high(more than $0.8$) when the model's accuracy is very low(less than $0.4$). 
This is due to its failure to consider whether the model is accurately locating the object's position at this time.
Secondly, the difference in classification confidence is smaller at the lowest and highest detection accuracy.
This makes it difficult to determine a suitable threshold to distinguish the quality of the detection results.
Finally, the localization confidence correlates lowest with the accuracy of the model.
It is because high localization confidence only ensures that the model accurately locates objects in the video frame.
However, the model may not accurately classify these objects due to the influence of data drift in real-world mobile scenes. 
In fact, detection confidence CLC has plenty of methods to express with CC and LC. According to this experiment, we find that the multiplication of CC and LC is simple and can reflect the fluctuation of mobile DNNs' accuracy well. Therefore, we use the product of CC and LC to represent CLC.

\begin{table}[]
\centering
\scriptsize
\caption{Performance of three fixed sample rate strategies with the optimal sampling strategy in various data drifts.}
\label{exp:tb:622}
\begin{tabular}{|c|c|c|c|}
\hline
\textbf{\begin{tabular}[c]{@{}c@{}}Data drift type\end{tabular}} & \textbf{\begin{tabular}[c]{@{}c@{}}Sampling strategy\end{tabular}} & \textbf{mAP(IoU=0.50)} & \textbf{\begin{tabular}[c]{@{}c@{}}$\mathbf{T_{infer}}$~\textbf{(s)}\end{tabular}} \\ \hline
\multirow{4}{*}{\textbf{Sudden}} & $[{t}_{1},{t}_{3}]\times 0.3$~fps & 0.545 & 305.6 \\ \cline{2-4} 
 & $[{t}_{1},{t}_{3}]\times 0.6$~fps & 0.652 & 348.2 \\ \cline{2-4} 
 & $[{t}_{1},{t}_{3}]\times 0.9$~fps & 0.587 & 283.9 \\ \cline{2-4} 
 & Optimal & 0.656 & 363.4 \\ \hline
\multirow{4}{*}{\textbf{Incremental}} & $[{t}_{1},{t}_{3}]\times 0.3$~fps & 0.510 & 208.4 \\ \cline{2-4} 
 & $[{t}_{1},{t}_{3}]\times 0.6$~fps & 0.576 & 267.9 \\ \cline{2-4} 
 & $[{t}_{1},{t}_{3}]\times 0.9$~fps & 0.474 & 167.4 \\ \cline{2-4} 
 & Optimal & 0.690 & 402.3 \\ \hline
\multirow{4}{*}{\textbf{Gradual}} & $[{t}_{1},{t}_{3}]\times 0.3$~fps & 0.506 & 128.6 \\ \cline{2-4} 
 & $[{t}_{1},{t}_{3}]\times 0.6$~fps & 0.510 & 173.9 \\ \cline{2-4} 
 & $[{t}_{1},{t}_{3}]\times 0.9$~fps & 0.394 & 101.9 \\ \cline{2-4} 
 & Optimal & 0.669 & 376.4 \\ \hline
\end{tabular}
\end{table}

\begin{table*}[]
\centering
\caption{System performance of multiple model evolution tasks in terms of inference accuracy and evolving time.}
\vspace{-0.1cm}
\scriptsize
\scalebox{1}{
\begin{tabular}{|c|c|c|c|cc|cc|ccc|}
\hline
\multirow{2}{*}{\textbf{\begin{tabular}[c]{@{}c@{}}Edge\\ conf.\end{tabular}}} & \multirow{2}{*}{\textbf{\begin{tabular}[c]{@{}c@{}}Scheduling\\ method\end{tabular}}} & \multirow{2}{*}{\textbf{\begin{tabular}[c]{@{}c@{}}Task \\ number\end{tabular}}} & \multirow{2}{*}{\textbf{\begin{tabular}[c]{@{}c@{}}Avg evolving \\ time(s)\end{tabular}}} & \multicolumn{2}{c|}{\textbf{\begin{tabular}[c]{@{}c@{}}Scheduling \\ time(s)\end{tabular}}} & \multicolumn{2}{c|}{\textbf{\begin{tabular}[c]{@{}c@{}}Retraining \\ time(s)\end{tabular}}} & \multicolumn{3}{c|}{\textbf{mAP}} \\ \cline{5-11}
&  &  &  & \multicolumn{1}{c|}{\textbf{Avg}} & \textbf{SD} & \multicolumn{1}{c|}{\textbf{Avg}} & \textbf{SD} & \multicolumn{1}{c|}{\textbf{IoU=0.50:0.05:0.95}} & \multicolumn{1}{c|}{\textbf{IoU=0.50}} & \textbf{IoU=0.75} \\ \hline
\multirow{4}{*}{\textbf{No. 1}} & \textbf{Default} & 4 & 146.4 & \multicolumn{1}{c|}{98.8} & 98.8 & \multicolumn{1}{c|}{46.0} & 19.0 & \multicolumn{1}{c|}{0.428} & \multicolumn{1}{c|}{0.587} & 0.494 \\ \cline{2-11}
& \textbf{Task scheduling} & 4 & 57.9 & \multicolumn{1}{c|}{8.1} & 7.5 & \multicolumn{1}{c|}{26.5} & 18.3 & \multicolumn{1}{c|}{0.518} & \multicolumn{1}{c|}{0.621} & 0.563 \\ \cline{2-11}
& \textbf{Resource allocation} & 4 & 59.9 & \multicolumn{1}{c|}{8.6} & 7.9 & \multicolumn{1}{c|}{27.8} & 11.0 & \multicolumn{1}{c|}{0.515} & \multicolumn{1}{c|}{0.618} & 0.559 \\ \cline{2-11}
& \textbf{Task scheduling + resource allocation} & 4 & 54.7 & \multicolumn{1}{c|}{6.3} & 5.2  & \multicolumn{1}{c|}{25.8} & 15.1 & \multicolumn{1}{c|}{0.589} & \multicolumn{1}{c|}{0.704} & 0.633 \\ \hline
\multirow{4}{*}{\textbf{No. 2}} & \textbf{Default} & 8 & 148.6 & \multicolumn{1}{c|}{99.1} & 99.1 & \multicolumn{1}{c|}{49.3} & 27.4 & \multicolumn{1}{c|}{0.417} & \multicolumn{1}{c|}{0.556} & 0.483 \\ \cline{2-11}
& \textbf{Task scheduling} & 8 & 57.8 & \multicolumn{1}{c|}{8.4} & 7.4 & \multicolumn{1}{c|}{25.6} & 18.7 & \multicolumn{1}{c|}{0.488} & \multicolumn{1}{c|}{0.591} & 0.533 \\ \cline{2-11}
& \textbf{Resource allocation} & 8 & 67.1 & \multicolumn{1}{c|}{15.4} & 17.7 & \multicolumn{1}{c|}{28.3} & 12.3 & \multicolumn{1}{c|}{0.495} & \multicolumn{1}{c|}{0.604} & 0.554 \\ \cline{2-11}
& \textbf{Task scheduling + resource allocation} & 8 & 54.6 & \multicolumn{1}{c|}{4.9} & 5.0 & \multicolumn{1}{c|}{26.2} & 14.9 & \multicolumn{1}{c|}{0.579} & \multicolumn{1}{c|}{0.679} & 0.611 \\ \hline
\end{tabular}
}
\label{tb_motivation_edge}
\vspace{-0.3cm}
\end{table*}

\subsubsection{Performance of Fixed Sampling Rate Strategy in Three Diverse Data Drifts}\label{exp:212}

To illustrate the need for the adaptive frame sampling strategy (as discussed in \S \ref{sec:43}),
we compare the performance of three fixed sample rate strategies (\ie0.3~fps, 0.6~fps, 0.9~fps) with the optimal sampling strategy in various data drifts from real-world mobile scenes.
The optimal sampling strategy is to go offline to select those video frames whose model inference results deviate from the ground truth.
Table~\ref{exp:tb:622} shows the accuracy(mAP(IoU=0.50)) and $T_{infer}$ corresponding to each sampling strategy in different data drifts.
We can see that for the sudden drift, the accuracy and $T_{infer}$ of the fixed sampling rate strategy(0.6~fps) are closest to the optimal sampling strategy, differing by only $0.4\%$ and $15.2s$, respectively.
As for the other two data drift types, however, the fixed sampling rate strategy loses more than $10\%$ in the accuracy and shortens the $T_{infer}$ by nearly $2 \times$.
This is due to the short drift duration and uniform data distribution of the sudden drift. The fixed sampling rate strategy can ensure that redundant frames are removed while the video frames that best reflect the real-world mobile scene are selected.
Moreover, the fixed sampling strategy is simpler to execute and can quickly select video frames for retraining to compensate for the accuracy loss caused by the sudden drift promptly.
However, the other two data drift types are characterized by a more uneven data distribution during the drift conversion process.
The fixed sampling rate strategy leads to losing the best video frames. It makes selected video frames not fully reflect the characteristics of the new scene, resulting in the poor performance of the evolved model.

\subsubsection{Performance for Multiple Model Evolution Tasks }\label{exp:212}
As the number of mobile ends served by the server increases, the server's computing and memory resources can easily become the bottleneck of improving the system performance, resulting in long delays in DNN evolutions. 
The mobile end suffers more inference degradation without a timely evolved deep model. We conduct an experiment to demonstrate the system performance (inference accuracy and evolution latency)
when the edge servers handle multiple simultaneous evolution requests.
As shown in Table \ref{tb_motivation_edge}, the evolution latency is high, 
\eg $\geq 148.6s$, using the default GPU parallel scheduling algorithm that works in a first-come-first-serve
manner.  
This is because more evolution tasks that exceed the available GPU memory supply will keep waiting.
Our insights on such bottlenecks are:
\textit{(i)} the GPU memory resource imposes a \textit{hard threshold} on the execution of a re-training task,
\ie a task needs to acquire enough memory before execution. 
Thus, a task scheduling mechanism is necessary to arrange valuable memory resources for carefully-selected tasks, 
to maximize the average QoE for all mobile ends.
\textit{(ii)} the GPU computation resource is a \textit{tunable variable} that can be dynamically adjusted to 
speed up an arbitrary evolution task and thereby shorten the retraining time.
As shown in Table \ref{tb_motivation_edge}, the cooperative task scheduling and resource allocation schemes reduce the average scheduling 
and retraining time by up to $20.2 \times$ and $1.9 \times$ over two diverse server configurations. 
In summary, the \textit{bottleneck} that limits the model retraining latency is the edge server's GPU resources.

\subsection{Performance Comparison}
\label{subsec:exp_62}

\subsubsection{Performance Comparison for Single-mobile DNN Case}
\label{subsec:exp_621}
This experiment compares the performance of \sysname and three different baseline methods on real-world video clips.
We compare the \sysname with baselines(\ie A1, A2, A3) on four real-world mobile videos (D1$\sim$ D4) with a network bandwidth of 10.65 MB/s. These videos are all thirty minutes captured by onboard cameras. The mobile ends under different methods are all deployed with the model1 after 8-bit quantization. We run the measurements and report the life-cycle accuracy of models retrained by \sysname and three other baseline methods in each video.
\sysname achieves the best inference accuracy compared to the original compressed model (A1), the domain adaptation (A2), and the down-sampling method baseline.
Table \ref{tb_621} shows the mean average precision (mAP) of A1, A2, A3, and \sysname under different intersections over union (IoU) thresholds.
Compared with the original model (A1), \sysname improves mAP by $22.9\%$, $34\%$, and $29.3\%$ at different IoU thresholds (IoU=0.50:0.05:0.95, IoU=0.50, IoU=0.75), respectively. 
Compared with the domain adaptation method (A2), mAP is improved by $13.6\%$, $20.4\%$, and $25.5\%$, respectively, with three IoU thresholds.
Compared with the down-sampling method (A3), mAP is improved by $14.9\%$, $20.9\%$, and $19.4\%$, respectively, with three IoU thresholds.
\begin{table}[]
\scriptsize
\caption{Accuracy comparison for the single mobile end.}
\begin{tabular}{|c|c|c|c|c|c|}
\hline
\multicolumn{1}{|c|}{\begin{tabular}[c]{@{}c@{}}\textbf{Mobile collected}\\ \textbf{videos}\end{tabular}} & \textbf{mAP} & \textbf{A1} & \textbf{A2} & \textbf{A3} & \textbf{AdaEvo} \\ \hline
\multirow{3}{*}{\textbf{D1}} & \textbf{IoU=0.50:0.05:0.95} & 0.266 & 0.366 & 0.361  & 0.507 \\ \cline{2-6} 
                             & \textbf{IoU=0.50}           & 0.387 & 0.519 & 0.521  & 0.71  \\ \cline{2-6} 
                             & \textbf{IoU=0.75}           & 0.263 & 0.337 & 0.423 & 0.608 \\ \hline
\multirow{3}{*}{\textbf{D2}} & \textbf{IoU=0.50:0.05:0.95} & 0.312 & 0.412 & 0.373  & 0.496 \\ \cline{2-6} 
                             & \textbf{IoU=0.50}           & 0.414 & 0.532 & 0.536  & 0.709 \\ \cline{2-6} 
                             & \textbf{IoU=0.75}             & 0.354 & 0.425 & 0.436  & 0.585 \\ \hline
\multirow{3}{*}{\textbf{D3}} & \textbf{IoU=0.50:0.05:0.95} & 0.253 & 0.345 & 0.340  & 0.503 \\ \cline{2-6} 
                             & \textbf{IoU=0.50}          & 0.372 & 0.503 & 0.488  & 0.725 \\ \cline{2-6} 
                             & \textbf{IoU=0.75}             & 0.276 & 0.364 & 0.358  & 0.593 \\ \hline
\multirow{3}{*}{\textbf{D4}} & \textbf{IoU=0.50:0.05:0.95} & 0.223 & 0.304 & 0.3    & 0.464 \\ \cline{2-6} 
                             & \textbf{IoU=0.50}           & 0.315 & 0.478 & 0.469  & 0.705 \\ \cline{2-6} 
                             & \textbf{IoU=0.75}             & 0.263 & 0.345 & 0.333  & 0.54  \\ \hline
\end{tabular}
\label{tb_621}
\end{table}

\begin{table*}[]
\scriptsize
\caption{Performance comparison between cloud- and edge-assisted schemes.}
\begin{tabular}{|c|c|c|ccc|c|c|l|c|ccc|}
\hline
\multirow{2}{*}{} & \multirow{2}{*}{\begin{tabular}[c]{@{}c@{}}\textbf{Uplink} \\ \textbf{(MB/s)}\end{tabular}} & \multirow{2}{*}{\begin{tabular}[c]{@{}c@{}}\textbf{Downlink} \\ \textbf{(MB/s)}\end{tabular}} & \multicolumn{3}{c|}{\textbf{Evolving time(s)}} & \multirow{2}{*}{\begin{tabular}[c]{@{}c@{}}$\mathbf{{T}_{retrain}}$\\ \textbf{(s)}\end{tabular}} & \multirow{2}{*}{\begin{tabular}[c]{@{}c@{}}{$\mathbf{{T}_{infer}}$}\\ \textbf{(s)}\end{tabular}} & \multicolumn{1}{c|}{\multirow{2}{*}{\textbf{Q}}} & \multirow{2}{*}{$\mathbf{{R}_{t}}$} & \multicolumn{3}{c|}{\textbf{mAP}} \\ \cline{4-6} \cline{11-13} 
 &  &  & \multicolumn{1}{c|}{$\mathbf{{t}_{u}}$} & \multicolumn{1}{c|}{$\mathbf{{t}_{r}}$} & $\mathbf{{t}_{d}}$ &  &  & \multicolumn{1}{c|}{} &  & \multicolumn{1}{c|}{\textbf{IoU=0.50:0.05:0.95}} & \multicolumn{1}{c|}{\textbf{IoU=0.50}} & \textbf{IoU=0.75} \\ \hline
\textbf{Cloud-assisted A4} & 0.85 & 0.67 & \multicolumn{1}{c|}{51.1} & \multicolumn{1}{c|}{20.5} & 315.8 & 387.4 & 52.5 & 228.3 & 11.9\% & \multicolumn{1}{c|}{0.389} & \multicolumn{1}{c|}{0.519} & 0.454 \\ \hline
\textbf{Edge-assisted AdaEvo} & 11.25 & 10.61 & \multicolumn{1}{c|}{3.9} & \multicolumn{1}{c|}{45.5} & 19.9 & 69.3 & 330.7 & 313.6 & 82.7\% & \multicolumn{1}{c|}{0.561} & \multicolumn{1}{c|}{0.784} & 0.674 \\ \hline
\end{tabular}
\label{tb_622}
\end{table*}
 
\subsubsection{Performance Comparison between Cloud Server- and Edge-assisted Schemes}
\label{subsec:exp_622}
This experiment compares the performance of \sysname with that of the cloud-based model evolution baseline (A4) for a single mobile end.
We leverage \sysname and cloud-based model evolution to test mobile ends deployed with model1 after 8-bit quantization on a mobile video (D3).
We introduce four retraining metrics, \ie total evolving time $T_{retrain}$ (Equ. (\ref{eqn:t_retrain})),time ratio $R_t$, accuracy, and QoE $Q_i$.
Table \ref{tb_622} summarizes the evaluation results on testing data (D3). 
%
First, compared with edge-assisted \sysname, the evolving time for the cloud-based scheme (A4) is significantly longer ($5.6 \times$) due to the low uplink and downlink bandwidths to the cloud. 
Second, the average accuracy of \sysname during the full life-cycle (\ie $T_{infer}+T_{retrain}$ formulated in \S \ref{sec:form}) is much higher than the cloud-assisted scheme (A4).
This is because the longer evolving time $T_{retrain}$ in A4 results in a long duration for the mobile end to endure the low-accuracy model. %
The longer the evolving time, the lower the average accuracy over the life cycle.
Third, \sysname improves the QoE $Q$ by $1.4 \times$, compared to A4. 







\subsubsection{Performance Comparison for Multi-mobile DNNs Case}
\label{subsec:exp_623}
\begin{figure}[t]
  \hfill
  \subfloat[Evolving time] {\includegraphics[width=0.21\textwidth]{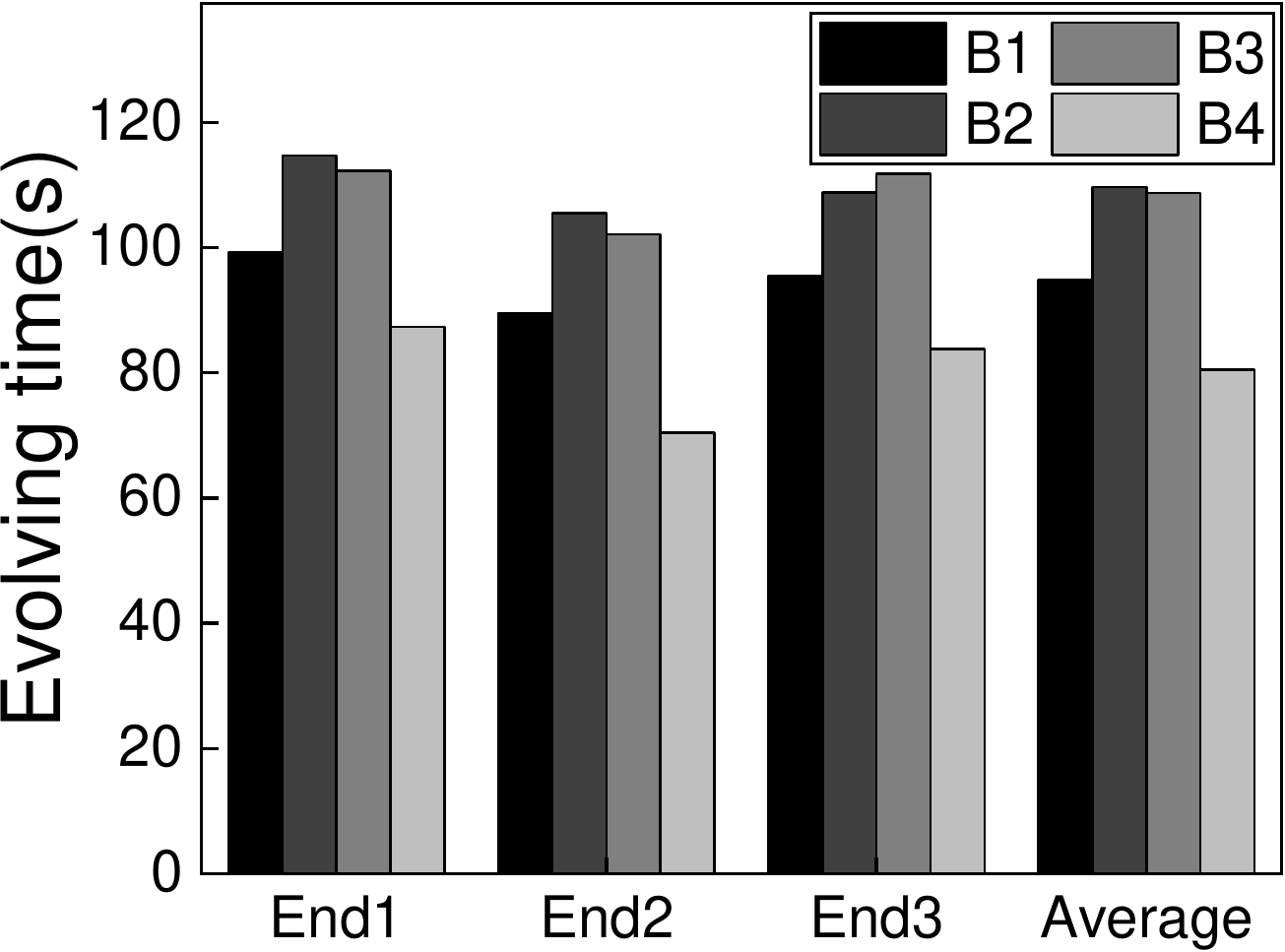}}
  \hfill
  \subfloat[QoE] {\includegraphics[width=0.21\textwidth]{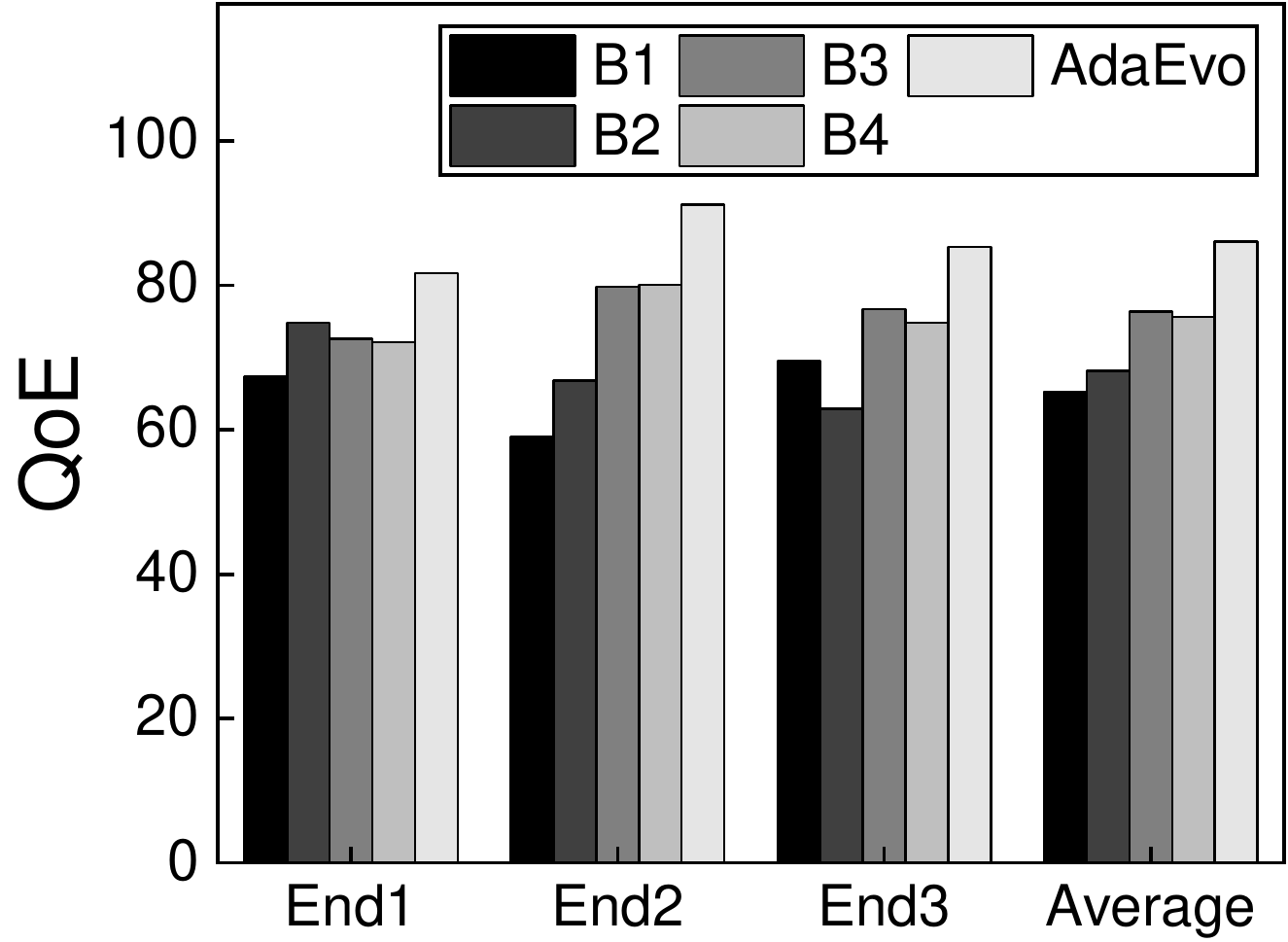}}
  \hfill
    \hspace{0.4cm}
  \caption{Performance comparison for multiple mobile ends.}
  \label{fig:ex_623}
 \end{figure}

This experiment compares the performance of \sysname and four baseline methods in the evolving time and QoE for three mobile ends which can send multiple evolution requests.
For the case where an edge server serves three mobile ends, as for evolving time, test the performance of the dynamic programming-based evolution task selection algorithm with three baseline scheduling methods (B1 $\sim$ B3) with the network bandwidth of 10.76MB/s. Then we estimate the average QoE for mobile ends of \sysname and the other four methods(B1 $\sim$ B4) under the same network communication conditions. Each mobile end in these two experiments completes eighteen tasks.
Figure~\ref{fig:ex_623} shows the test results and Figure~(a) illustrates the comparison of evolution time, and Figure~(b) presents the evaluation of QoE. First, compared with the two serial execution(B2 and B3), our dynamic programming-based evolution task selection algorithm achieves the minimum evolution time. For example, the evolution time of the dynamic programming algorithm is reduced by $26.58\%$ and $25.94\%$ compared to B2 and B3, respectively. This is because \sysname maximizes the throughput of GPU and utilizes the principle of the short job first, significantly reducing the scheduling time and retraining time.
Second, the average QoE of the default GPU scheduling(B1) is extremely low(65.29). By default, multiple tasks will compete for the GPU memory resources. Only a few are executed, and others need to keep waiting, including ones with high model evolving urgency. Third, although the dynamic programming algorithm(B4) obtains the lowest evolution latency, \sysname increases the average QoE by $15\%$ and $32\%$ compared with B4 and B1, respectively. Dynamic programming-based evolution task selection algorithm uses the SJF principle and maximizes the GPU throughput but fails to consider the model evolving urgency $\lambda_{i}$ which is significant to optimize QoE.

\subsection{System Performance}
\label{subsec:exp_63}

\subsubsection{Adaptation to Different Network Bandwidths}
\label{subsec:exp_631}

\begin{figure}[t]
  \hfill
  \subfloat[mAP(IoU=0.50)] {\includegraphics[width=0.21\textwidth]{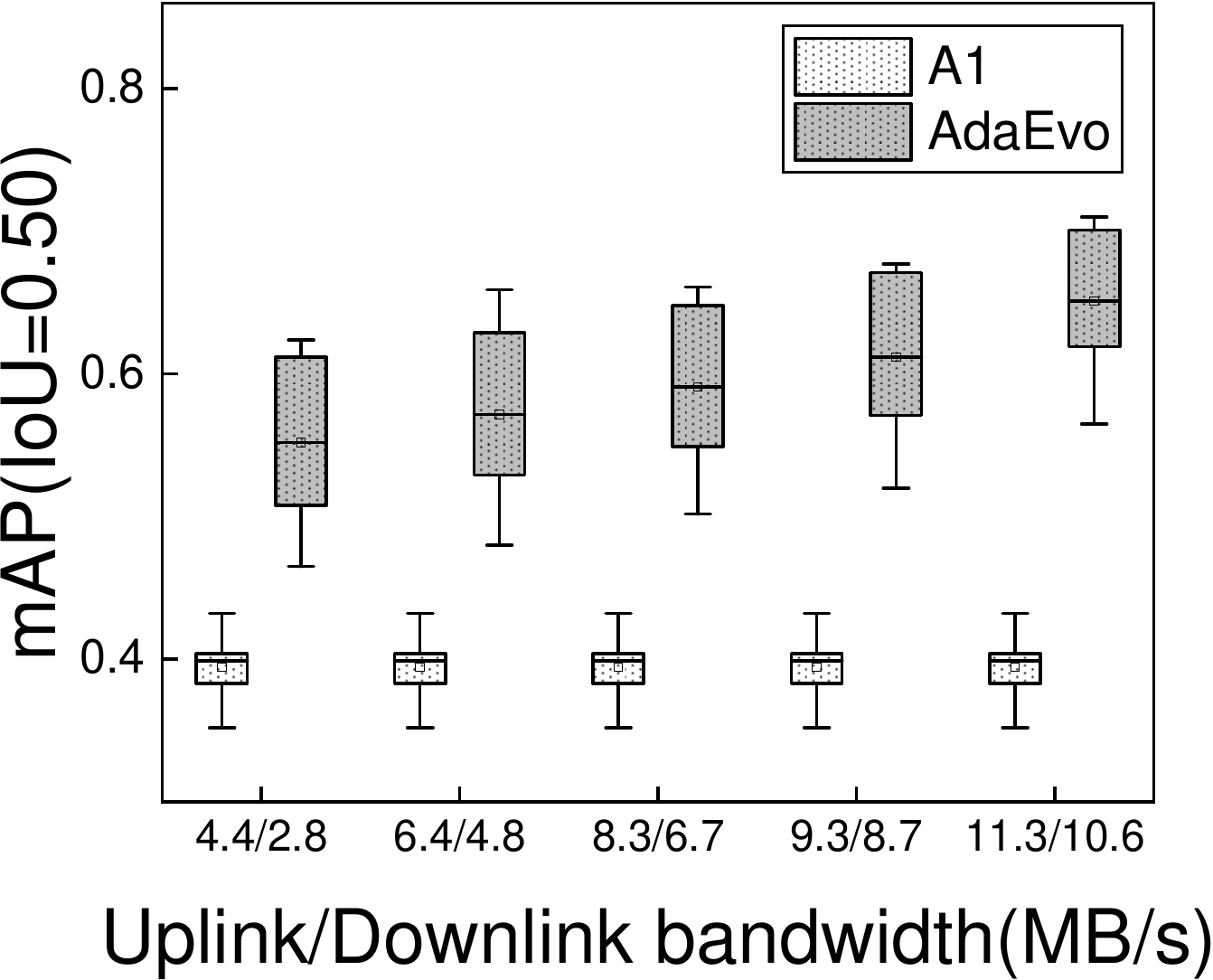}}
  \hfill
  \subfloat[Up/downloading time] {\includegraphics[width=0.21\textwidth]{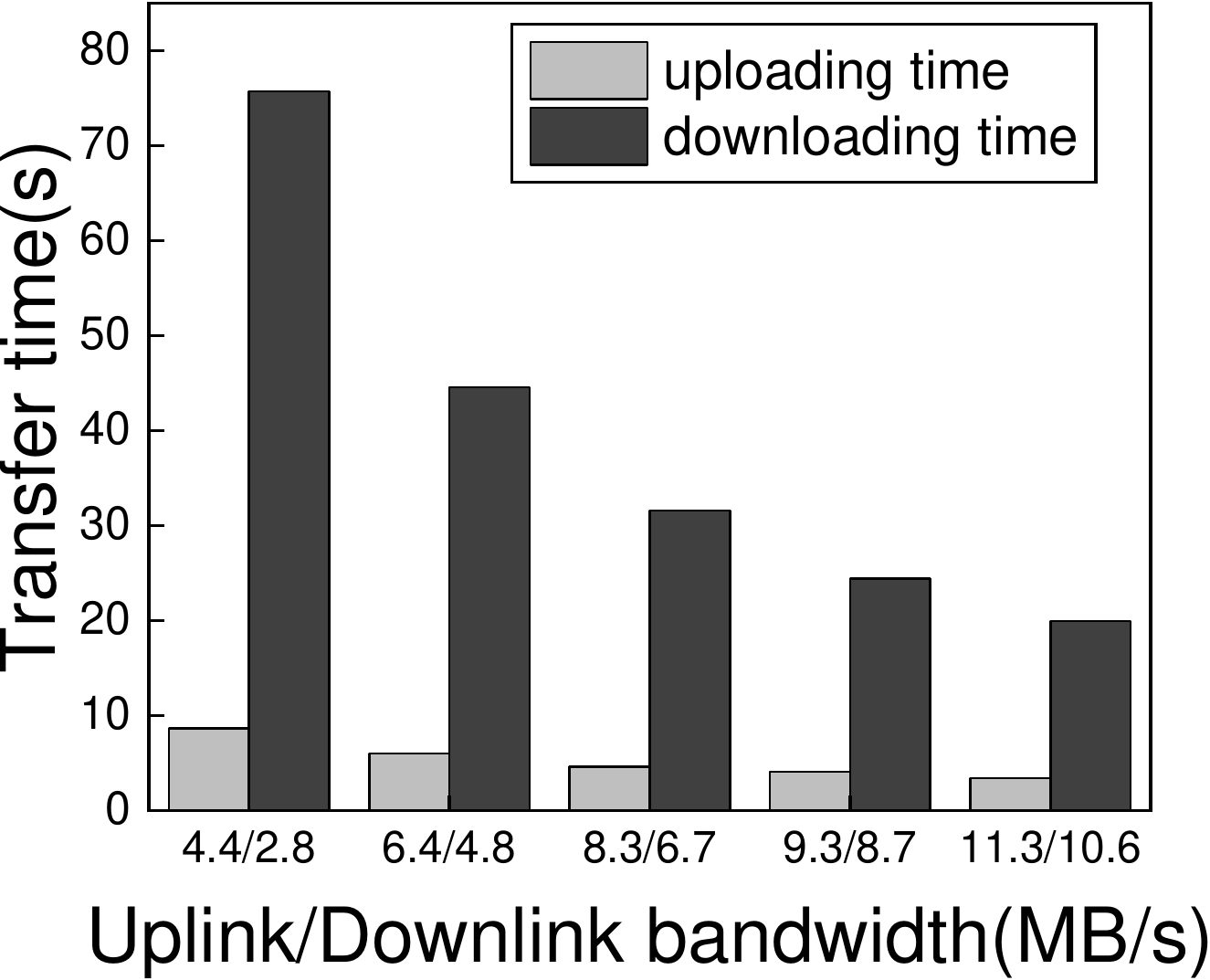}}
  \hspace{0.4cm}
  \caption{Performance with diverse link bandwidths.}
  \label{fig:ex_631}
\end{figure}

This experiment tests the \sysnameposs mAP and transmission time (\ie the uploading time ${t}_{u}$ and downloading time ${t}_{d}$) under five different network bandwidths, the results are shown in Figure~\ref{fig:ex_631}. 
%
%
First, with the decrease in bandwidth, the uploading time of video frames and the downloading time of the retrained parameters are getting longer, which increases the total evolving time, leading to degraded accuracy. 
Second, \sysname dramatically improves the accuracy by $15.8\% \sim 25.7\%$, compared with the original compressed model (A1) over different network bandwidths. 

\subsubsection{Robustness with Different Compression Budgets}
\label{subsec:exp_632}

\begin{figure}[t]
  \hfill
  \subfloat[Model quantification] {\includegraphics[width=0.2\textwidth]{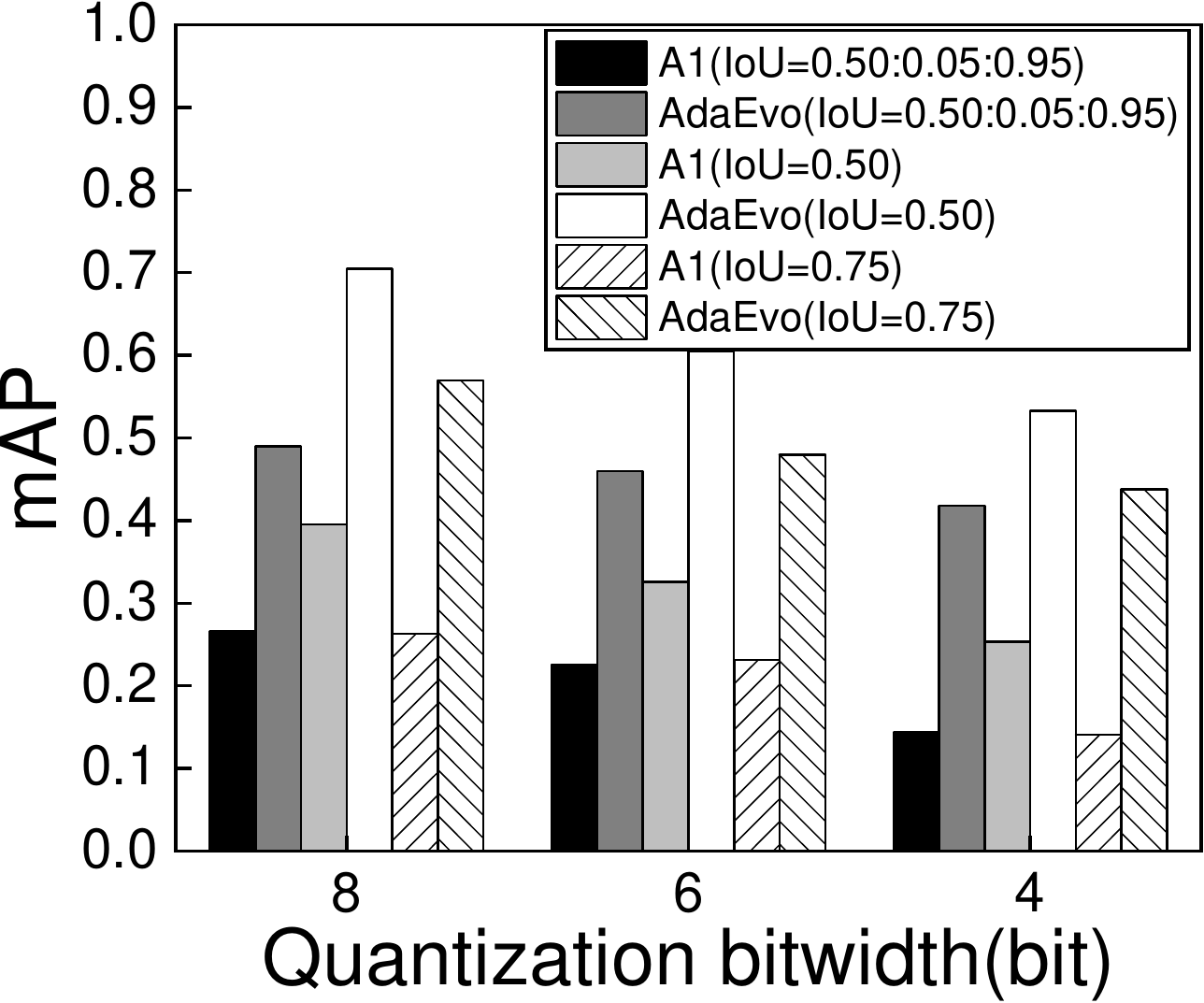}}
  \hfill
  \subfloat[Model pruning] {\includegraphics[width=0.2\textwidth]{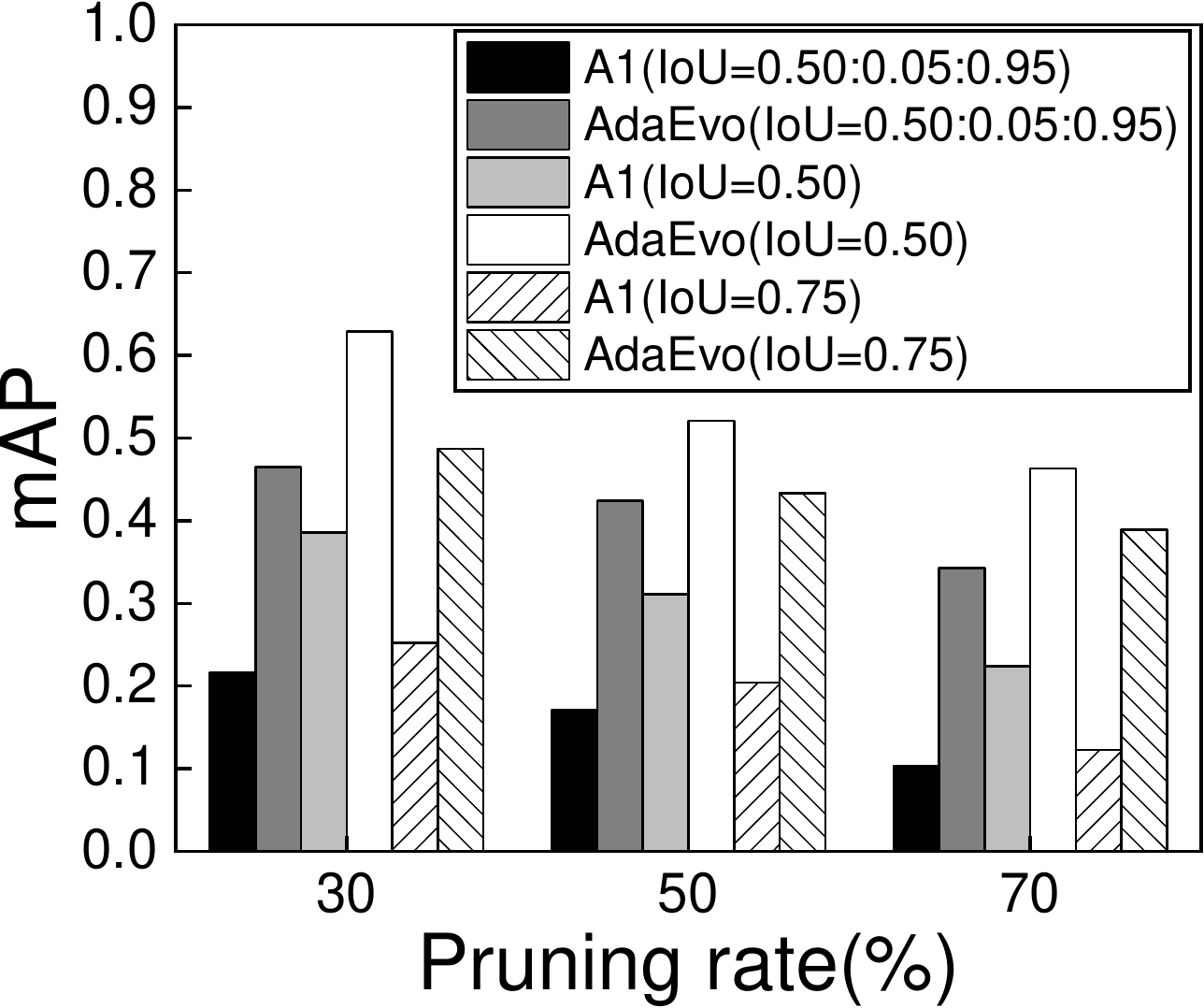}}
  \hspace{0.6cm}
  \caption{Performance on diverse compressed models.}
  \label{fig:ex_632}
\end{figure}

This experiment illustrates that \sysname can robustly calibrate the inference accuracy for diverse compressed deep models with various resource budgets imposed by mobile ends.
We test the performance of \sysname with six compressed variants of model 1 (see model details in \S \ref{sec_exp_setup}), \ie $8$-bit quantization, $6$-bit quantization, $4$-bit quantization, $30\%$ pruning, $50\%$ pruning, and $70\%$ pruning.
Figure \ref{fig:ex_632} shows \sysnameposs mAP for these different compression variants. 
\sysname can adaptively calibrate the detection accuracy for these diverse variants of compressed models.

\begin{figure}[htbp]
\centering
\begin{minipage}[t]{0.24\textwidth}
\centering
\includegraphics[width=3.6cm]{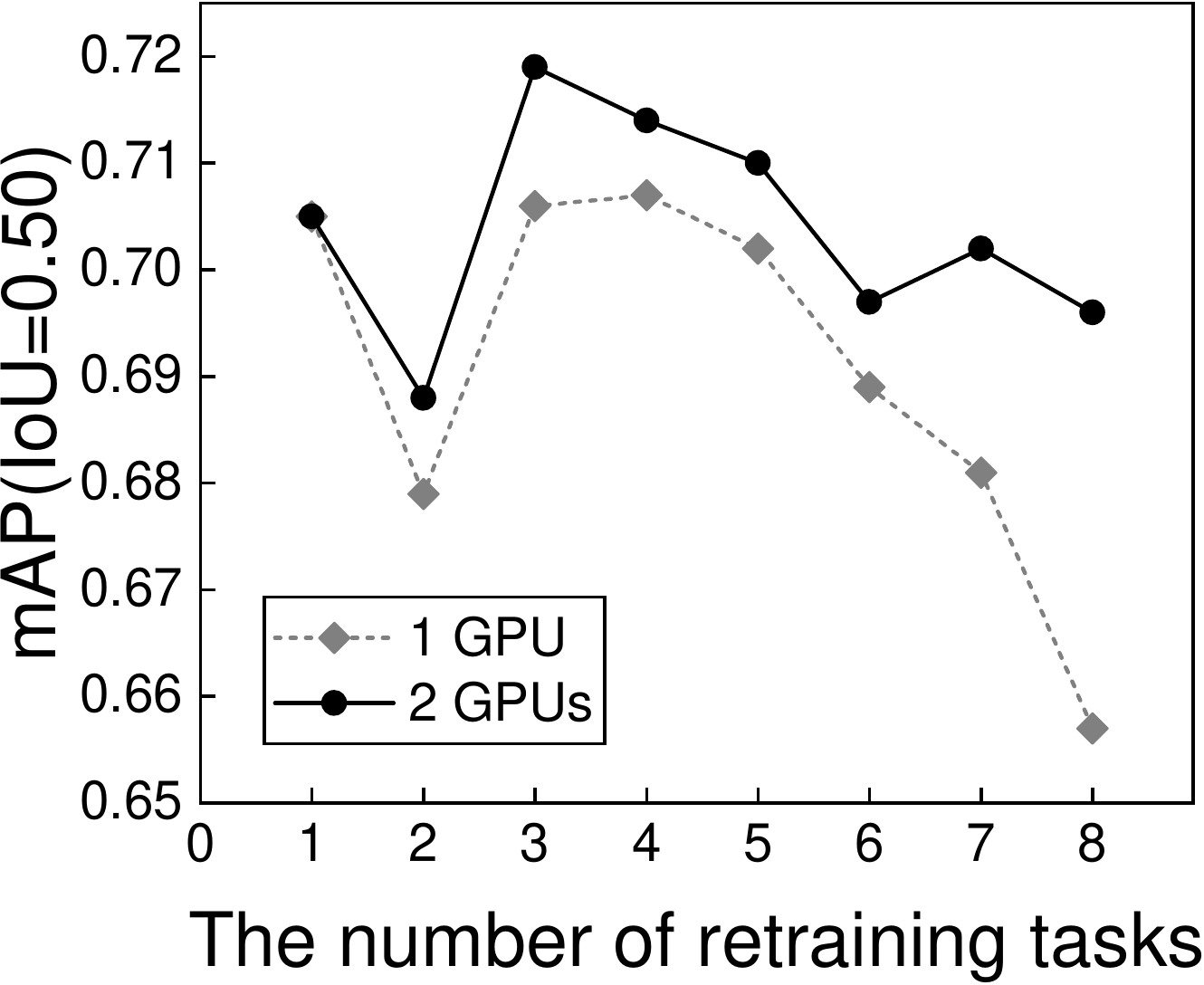}
		\caption{Performance on diverse edge servers.}
            \vspace{-0.28cm}
		\label{fig:ex_633}
\end{minipage}
\begin{minipage}[t]{0.24\textwidth}
\centering
\includegraphics[width=3.6cm]{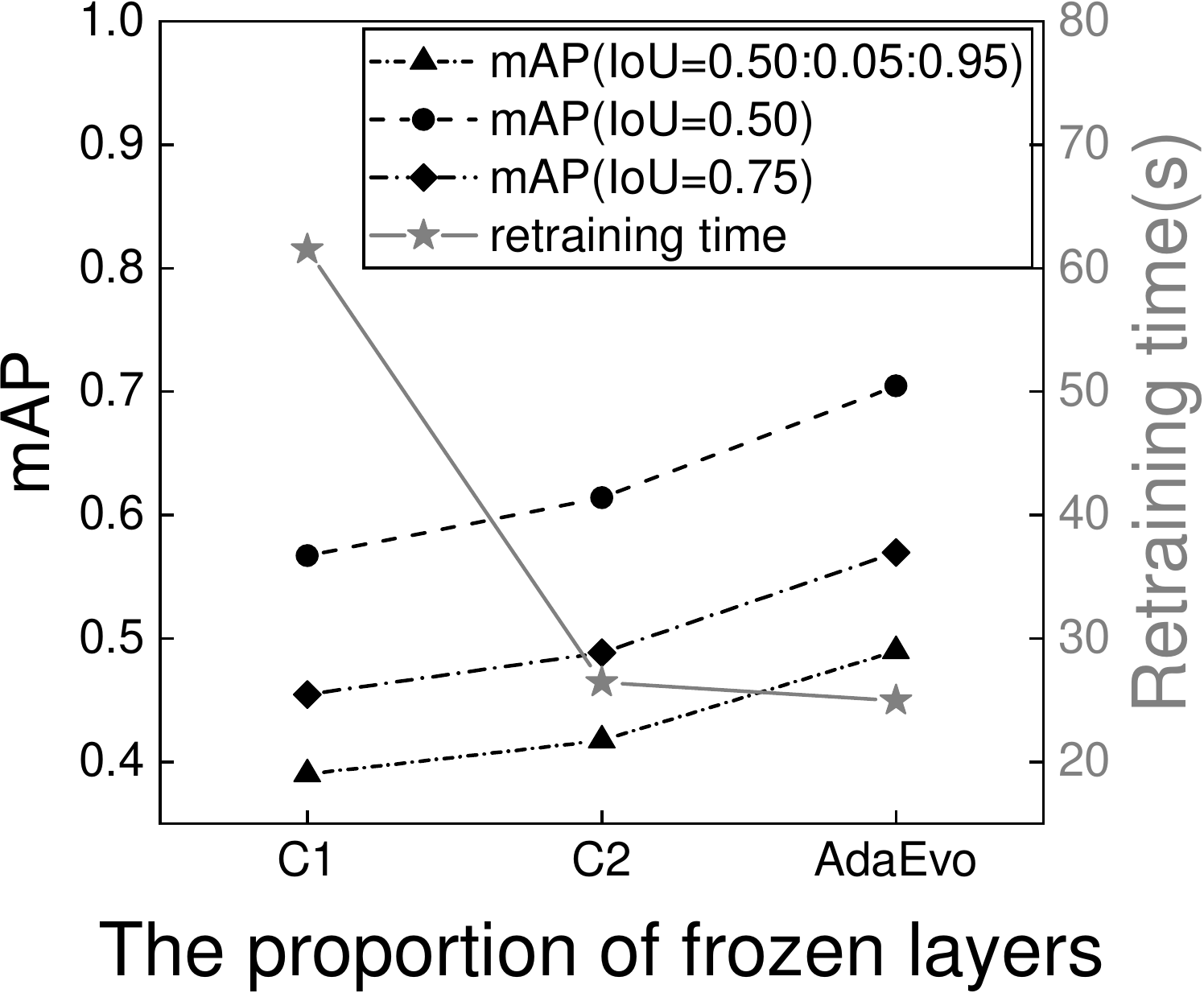}
		\caption{Impact of freeze retraining.}
            \vspace{-0.28cm}
		\label{fig:ex_643}
\end{minipage}
\end{figure}

        

\begin{table*}[]
\centering
\caption{Performance comparison of different evolution trigger strategies}
\begin{tabular}{|c|c|c|c|c|}
\hline
\textbf{Evolution trigger strategy} & \textbf{Time error rate} & \textbf{mAP(IoU=0.50)} & \textbf{Average evolution interval(s)} & \textbf{Whether to use other resources} \\ \hline
\textbf{A5} & 9.86\% & 0.698 & 260 & YES \\ \hline
\textbf{A6} & 26.21\% & 0.618 & 200 & NO \\ \hline
\textbf{A7} & 16.50\% & 0.662 & 247 & NO \\ \hline
\textbf{AdaEvo} & 4.37\% & 0.761 & 368 & NO \\ \hline
\end{tabular}
\label{tb_accuracydrop}
\end{table*}

\subsubsection{Performance over Diverse Edge Servers}
\label{subsec:exp_633}

This experiment tests \sysname on edge servers with different numbers of GPUs, \ie one and two GPUs. 
%
%
We employ different numbers of evolution tasks, \ie one $\sim$ eight, to test the \sysnameposs performance on both servers.
The results are shown in Figure \ref{fig:ex_633}.
First, \sysname can adaptively generate suitable task scheduling strategies for different numbers of evolution tasks according to different edge server settings.
Second, despite the \sysnameposs adaptive ability, when the number of evolution tasks exceeds a certain number (\eg six in this experiment), the mAP gap between these two edge server cases becomes larger.
This is because some tasks still cannot obtain GPU memory resources to perform retraining in the weak server with one GPU. This eventually delayed some mobile ends to calibrate accuracy in time. 

\subsection{ Micro-benchmarks and Ablation Studies}

\label{subsec:exp_64}




\subsubsection{Performance of Adaptive Evolution Trigger}

To show the performance of the evolution trigger strategy, we compare \sysname with other existing works.
The baselines we compare are the cloud-assisted trigger method (A5)~\cite{bib:iccv2019:Mullapudi}, in which the mobile end periodically sends frames to the server for testing and determining the evolution time based on the accuracy drop, the fixed evolving frequency (A6)~\cite{bib:iccv2021:Khani}, in which the mobile end send evolution request to the server with a fixed frequency. The adaptive evolving frequency based on historical data (A7)~\cite{bib:iccv2021:Khani}, in which the evolving frequency is determined by the degree of changes in historical data.
We test the different evolution trigger methods of \sysname and other three ways in previous work on the real-world mobile video (D$3$) with a network bandwidth of 10.53 MB/s.
We introduce four key metrics,\ie time error rate, which means the error between the evolution time-point confirmed by each evolution trigger policy and the actual accuracy drop point, accuracy(mAP) during this video, average evolution intervals, and whether to use other resources.
Table~\ref{tb_accuracydrop} summarizes the evaluation results.
First, compared with A6 and A7, \sysname shows more accurate predictions for the DNN evolution time, with a $21.84\%$ and $12.13\%$ decrease in time error rate, respectively.
The A5 can obtain the accurate accuracy drop point, but the transmission delay causes the evolution lag. 
Second, \sysname achieves the highest accuracy with the least number of evolutions.
\sysname improves the accuracy by $6.3\%$, $14.3\%$, and $9.9\%$, respectively, compared to the other three methods. 
Besides, AdaEvo's average evolution interval is the longest, which effectively reduces the edge server's resource usage.
Finally, compared to A3, \sysname enables accurate triggering of evolutions without needing other computing resources while achieving optimal evolution performance.



\subsubsection{Impact of Data Drift-aware Video Frame Sampling}\label{subsec:exp_641}

\begin{figure} 
\centering 
\subfloat[Mobile collected video D1]{\label{fig:subfig:a}
\includegraphics[width=0.4\linewidth]{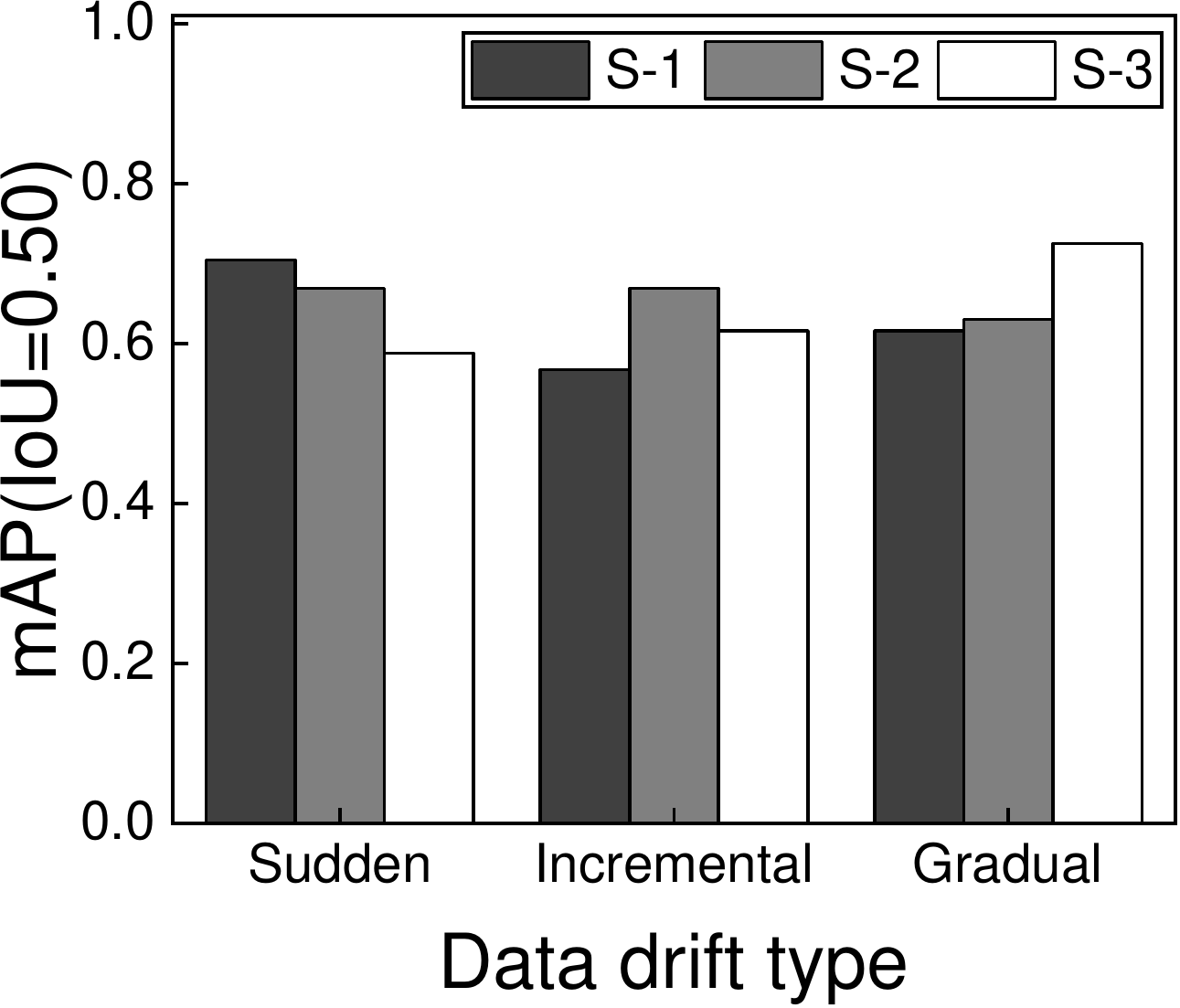}}
\hspace{0.01\linewidth}
\subfloat[Mobile collected video D2]{\label{fig:subfig:b}
\includegraphics[width=0.4\linewidth]{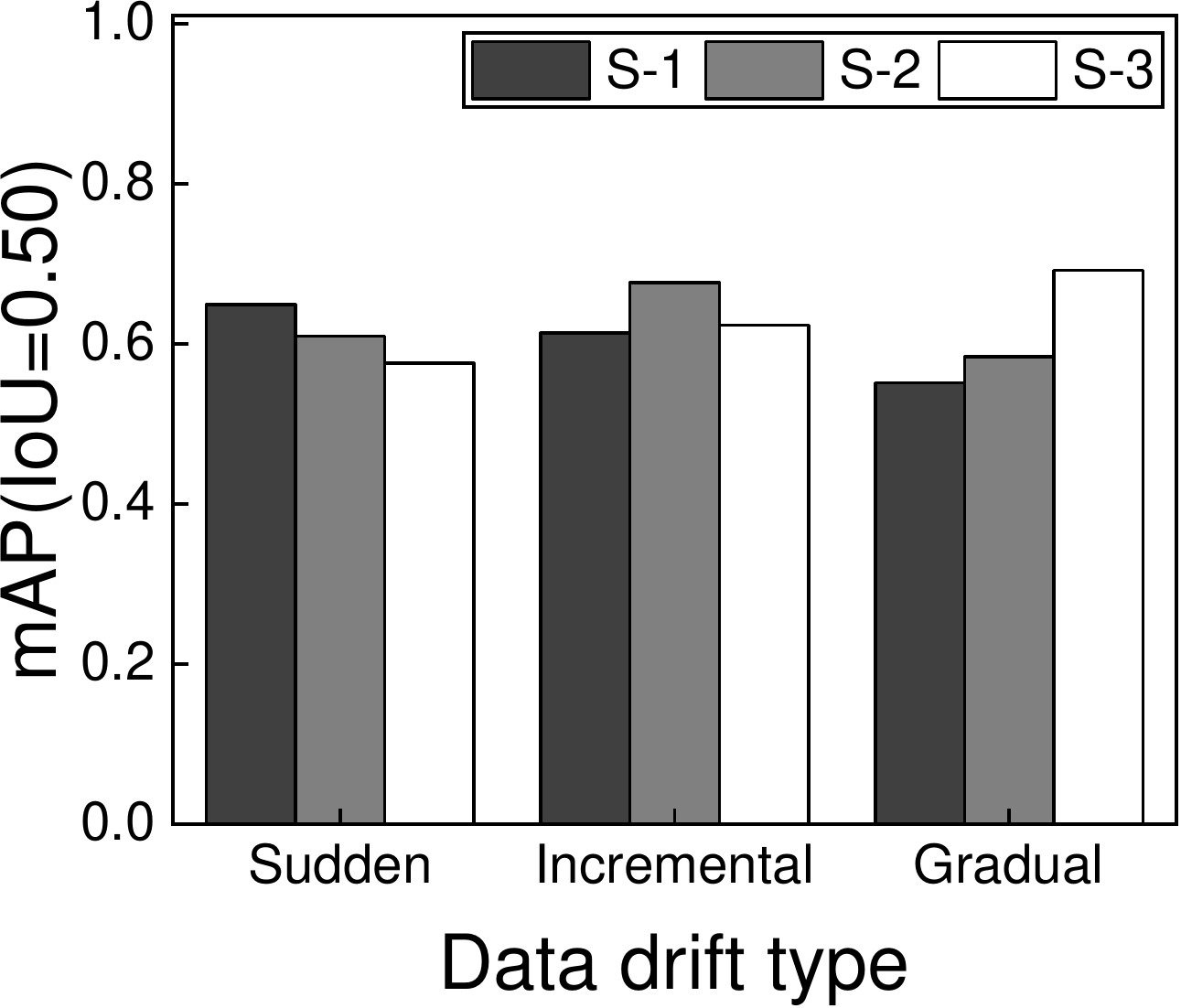}}
\vfill
\subfloat[Mobile collected video D3]{\label{fig:subfig:a}
\includegraphics[width=0.4\linewidth]{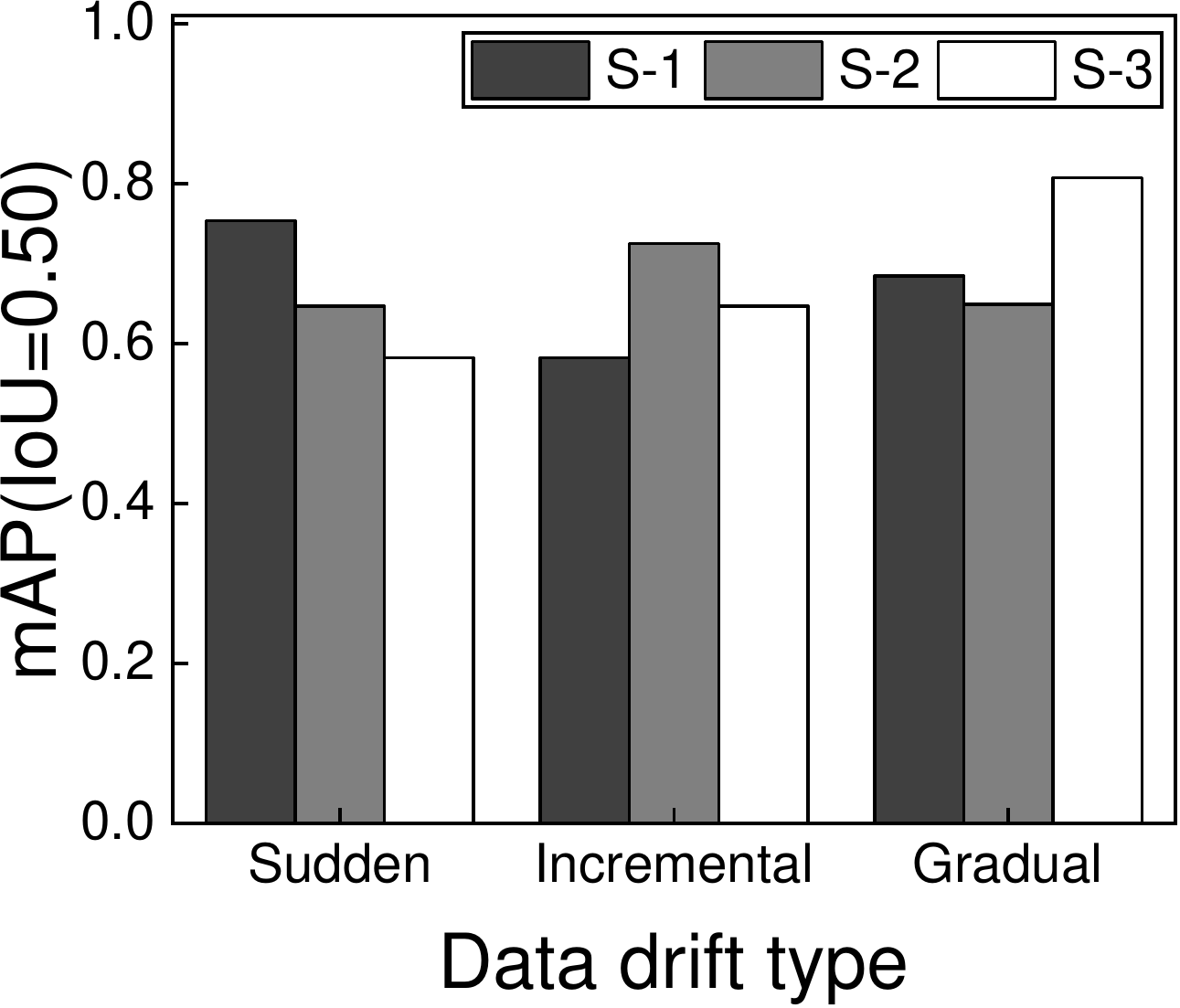}}
\hspace{0.01\linewidth}
\subfloat[Mobile collected video D4]{\label{fig:subfig:b}
\includegraphics[width=0.4\linewidth]{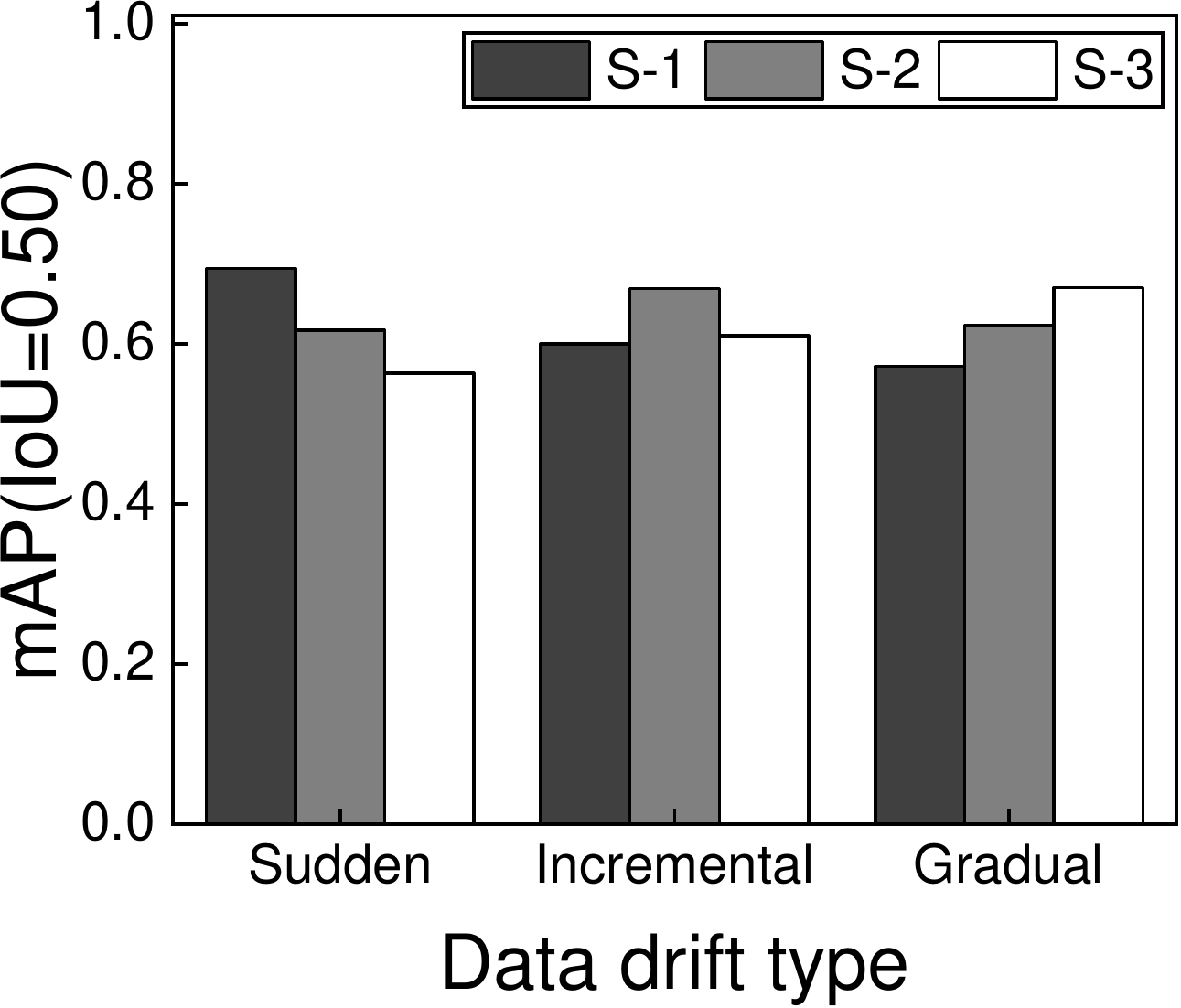}}
\caption{Performance comparison of different frame sampling strategies in multiple mobile scenes with three data drift types.}
\label{fig:641}
\end{figure}

We compare the detection accuracy of the three different sampling strategies in multiple mobile video datasets for different data drift types.
Figure~\ref{fig:641} illustrates the experimental results, where S-1, S-2, and S-3 denote the frame sampling strategies proposed by \sysname for sudden, incremental, and gradual drift, respectively.
We can see that the three frame sampling strategies achieve the best detection accuracy on their corresponding types of data drift.
First, the fixed frame sampling rate strategy (S-1) improves the detection accuracy by $9.1\%$ and $12.3\%$ in sudden drift compared to the other two strategies, respectively.
This is because the data transition process in sudden drift is shorter, and frames from the new scene are more evenly distributed after the drift is over. The fixed frame sampling rate strategy can reduce frame redundancy and quickly select useful frames to evolve the model in time to compensate for the accuracy loss.
Second, with the linear sampling rate strategy (S-2), the detection accuracy increases by $9.4\%$ and $6.6\%$ in incremental drift compared with the other two strategies.
This improvement can be attributed to the continuous transitions from old to new data in incremental drift. 
By adjusting the sampling rate based on it, we can capture video frames that accurately reflect the video characteristics of the new scene and enhances the evolution performance.
%
Finally, the frame-by-frame sampling strategy (S-3) improves the accuracy by $11.8\%$, $9.7\%$ in gradual drift.
Since the data distribution in the old and new scenes during gradual drift is more discrete. The frame-by-frame sampling strategy can accurately select frames that contribute more to retraining.

\subsubsection{Impact of Compression-aware DNN Freezing 
Retraining}
\label{subsec:exp_643}
To show the impact of compression-aware model freezing retraining (\S \ref{subsec:train_compress}), we compare \sysname with two retraining settings, \ie the full model retraining (C1) and randomly selected $31\%$ parameters for retraining(C2). 
As a separate note, we use $31\%$ to ensure the proportion of randomly selected parameters is similar to \sysname for a fair comparison.
Figure \ref{fig:ex_643} shows the experimental results. 
First, \sysname improves mAP(IoU=0.50) by $13.7\%$ compared to full model retraining. 
This is because \sysname finishes retraining faster (\eg $2.5 \times$), bringing accuracy gain sooner.
Second, compared to the scheme that randomly selects $70\%$ parameters to retrain, \sysname improves the performance by $9.0\%$ with the same retraining time.
It reflects that \sysnameposs retraining method improves the model generalization ability, thereby prolonging the high-accuracy inference duration.

\subsubsection{Impact of Grouping Strategy in Simplification of Search Space}
\label{subsec:exp_555}
We conduct an experiment to compare the different grouping strategies: equal urgency range grouping and equal task probability grouping. 
We take the total number of 12 mobile ends as an example and set the minimum number of tasks in each group $n_{min}$ and the length of urgency range threshold $\varepsilon$ as 3 and 35. Therefore, based on the adaptive group number decision, the suitable group number is four. Since the adaptive group number decision is designed based on equal task probability, we set more group numbers around four in this experiment.
Then, we use the normally distributed data characteristics to estimate the task probability and urgency range length of each group under the two different grouping strategies, with the number of groups being three, four, and five.
Table~\ref{tb:echelon} shows that equal urgency range grouping with group numbers of four and five obtain the extremely low task probability. In this case, only one or two tasks will likely appear in a group, and the scheduler cannot work well to select appropriate tasks from a global perspective. 
The maximum task probability of equal urgency range grouping with a group number of three is too high, which hurts the search speed of task selection. In addition, its average urgency range is wider than the equal task probability grouping with a group number of four, which will influence the optimization of $Q_{avg}$. Comprehensive considering, we employ the strategy of equal probability grouping by default in \sysname.

\subsubsection{Accuracy drop Rate in Detection of Evolution Trigger}
\label{subsec:exp_555}
The accuracy drop rate ($rod$) plays a crucial role in determining the detection of data drift and subsequently affects the frequency of model evolution. 
Figure~\ref{fig:rod} illustrates the average accuracy and the number of evolutions with mobile video (D1) under different $rod$ values.
When $rod$ is below 0.55, the average inference accuracy of the model experiences a significant drop. 
Conversely, when $rod$ exceeds 0.55, the number of model evolutions noticeably increases.
%
Therefore, we set $rod=0.55$ as the default value.
\subsubsection{Variance Threshold for Judging the End of Data Drift}
\label{subsec:exp_556}
To demonstrate $\alpha$ for judging the end of data drift, we compare the performance under different variance thresholds, as shown in Figure~\ref{fig:alpha}. We find that when $\alpha=(0.045)^{2}$, the accuracy of the evolved model reaches the highest. This is because when $\alpha$ is too high or too low, it will affect the judgment of data drift type, thereby affecting the strategy of video frame selection. Besides, $\alpha$ will affect the range of video frame selection. An inappropriate video frame selection range  will affect the selection of the most representative video frames, which hurts the model evolution effect.
\subsubsection{Thresholds for Gradual Drift}
\label{subsec:exp_557}
\begin{figure*}[htbp]
        
	\centering
	\begin{minipage}{0.32\linewidth}
		\centering
		\includegraphics[height=0.55\linewidth]{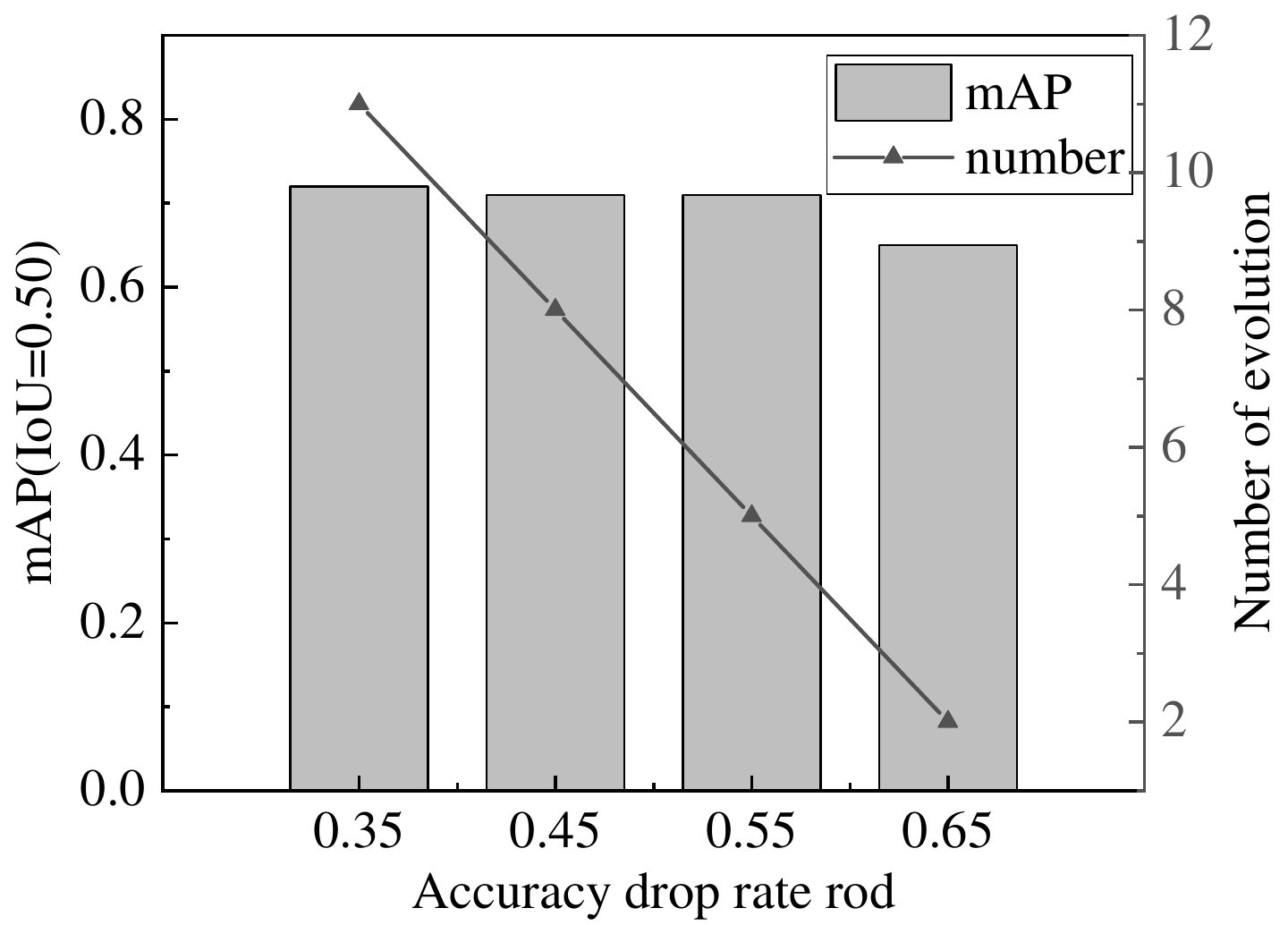}
		\caption{Performance on diverse  servers.}
		\label{fig:rod}
	\end{minipage}
        \hfill
	\begin{minipage}{0.32\linewidth}
		\centering
		\includegraphics[height=0.55\linewidth]{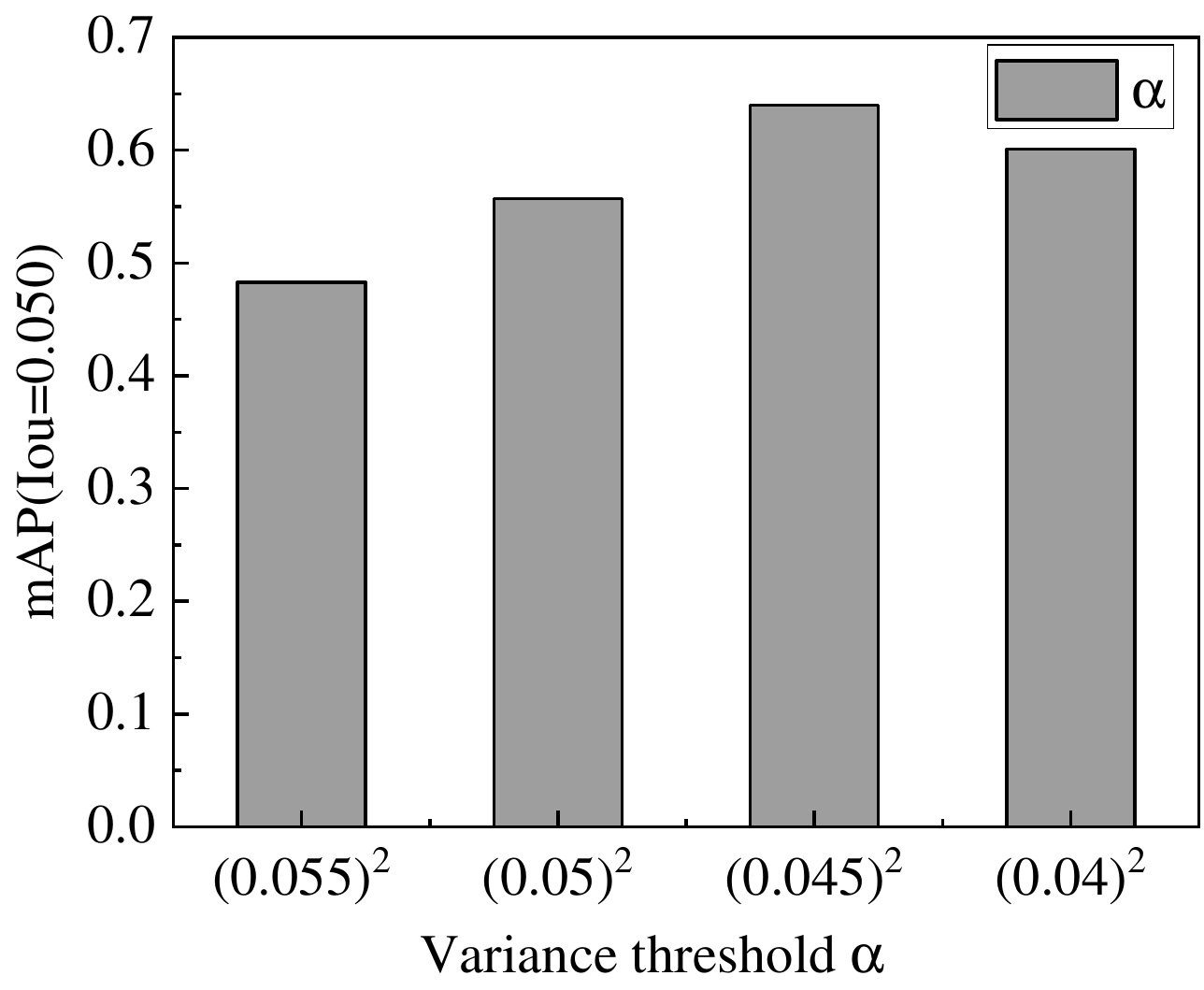}
		\caption{Impact of sample filter threshold.}
		\label{fig:alpha}
	\end{minipage}
        \hfill
	\begin{minipage}{0.32\linewidth}
		\centering
		\includegraphics[height=0.55\linewidth]{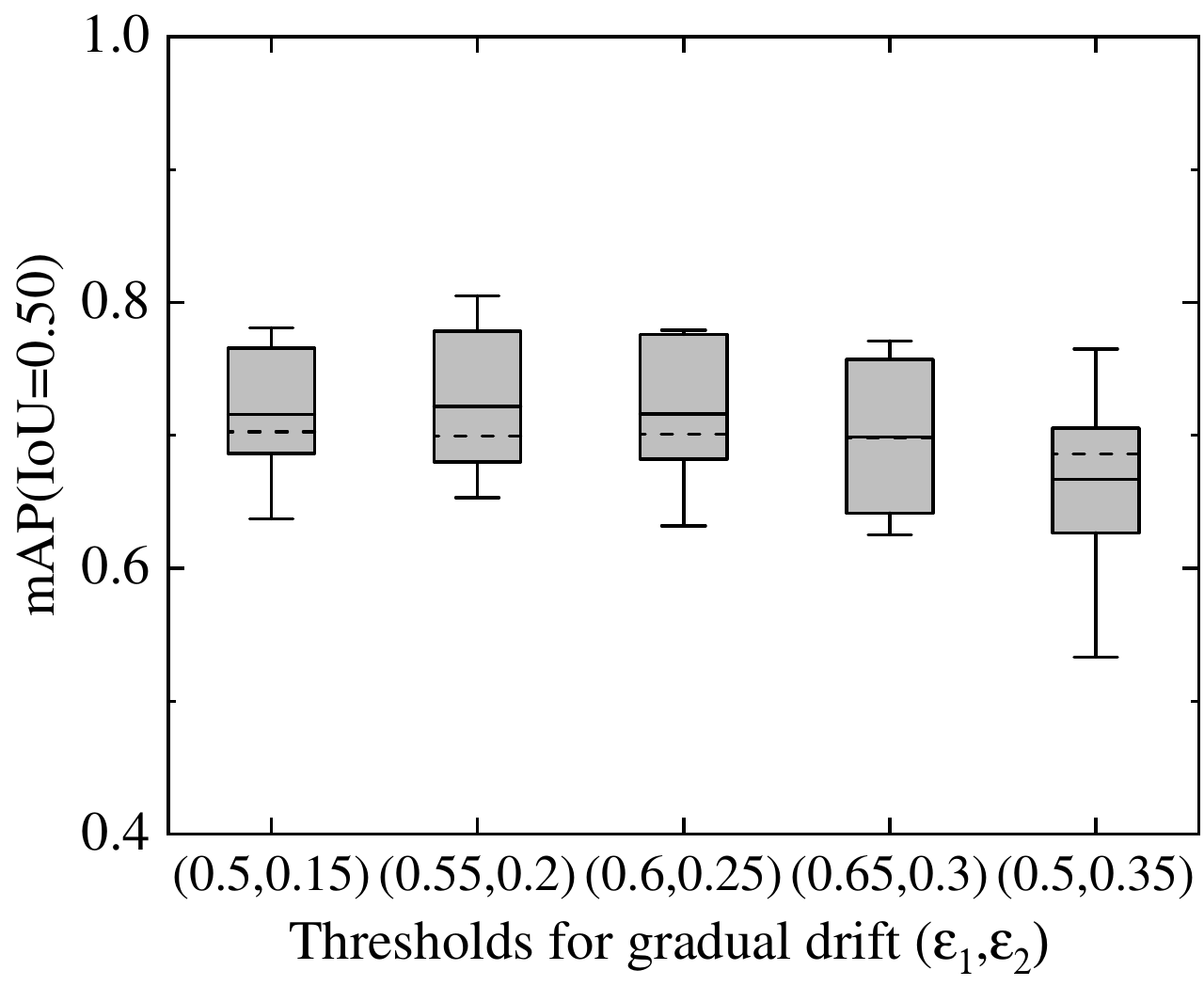}
		\caption{Impact of the threshold for gradual drift.}
		\label{fig:frame_difference}
	\end{minipage}
\end{figure*}

We conduct a comparison of the average accuracy over life-cycle in different scenarios, considering different values of $(\epsilon_{1},\epsilon_{2})$, as depicted in Figure~\ref{fig:frame_difference}. 
In our previous section (Section~\ref{sec:43}), we discussed the usage of $\epsilon_{1}$ in the frame difference method to alleviate the subsequent selection burden, while $\epsilon_{2}$ is employed for feature comparison to identify video frames that significantly contribute to the evolution and should be uploaded to the edge server.
We find that when the thresholds for gradual drift, $(\epsilon_{1},\epsilon_{2})$, were set to $(0.55, 0.2)$, model1 achieved the highest average inference accuracy.
%
Hence, finding an optimal balance for $\epsilon_{1}$ and $\epsilon_{2}$ is crucial to ensure an effective and efficient evolution.
\subsubsection{Acceptable Model Evolving Urgency Range Length in Task Grouping}
\label{subsec:exp_558}
As mentioned in \S~\ref{sec:searchspace}, the adjustment of ($\epsilon$) is mainly to control the upper limit of the number of tasks in each group. Therefore, $\epsilon$ has a close relationship with the number of groups. We conduct experiments to compare the performance under different $\epsilon$, where the number of evolution requests that the edge server can deal with concurrently($N$) is 12, as shown in Figure~\ref{fig:range_length}. 
In this case, when $\epsilon$ is 35, the number of groups is 4, and the average QoE of the task reaches the optimum. When $\epsilon$ takes other values, the change in the number of groups leads to changes in the search space, thereby hurting overall performance.
\subsubsection{Filtering Thresholds in Bounding Box-level Sample Filter}
\label{subsec:exp_642}
\begin{figure}[htbp]
\centering
\begin{minipage}[t]{0.24\textwidth}
\centering
\includegraphics[width=4.cm]{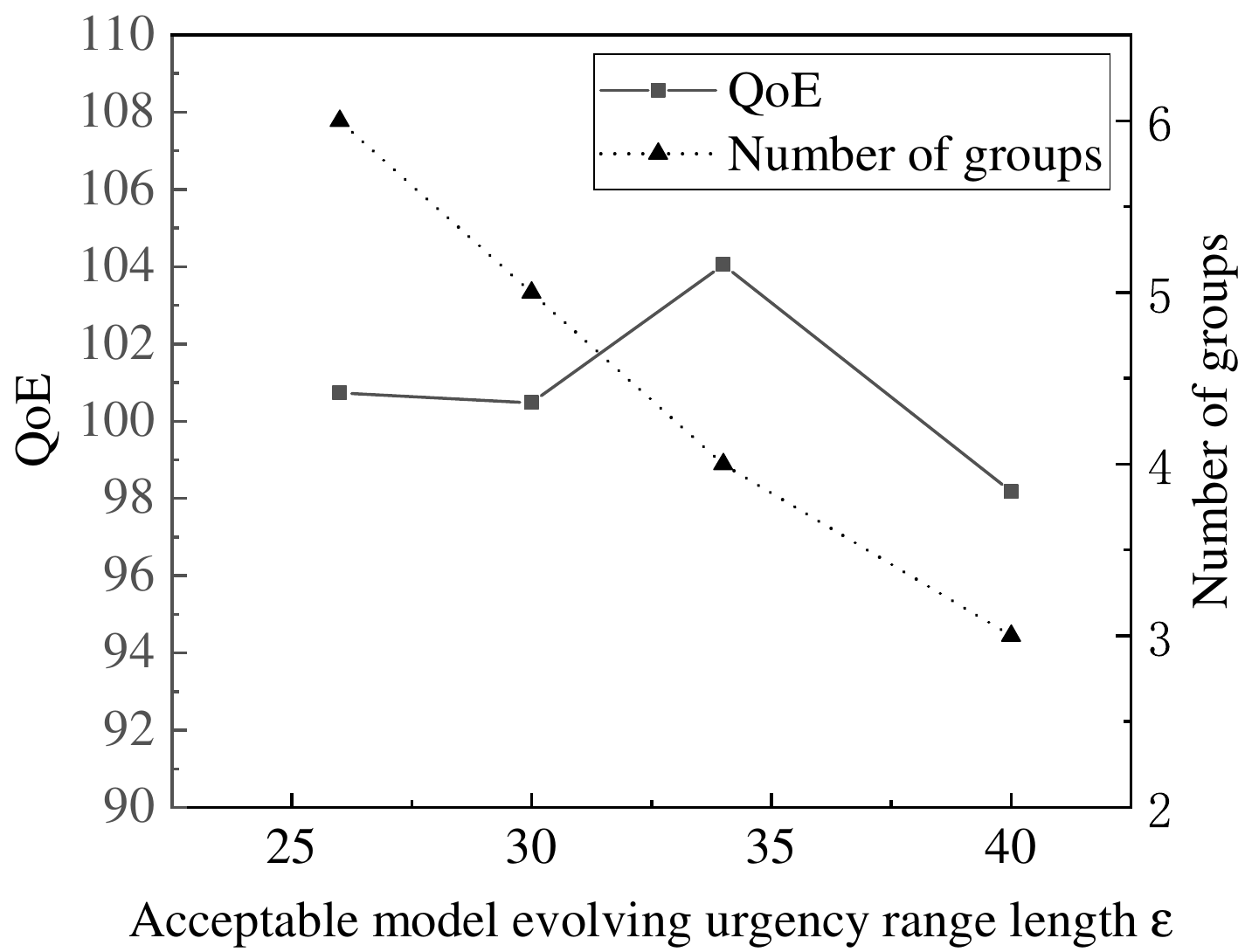}
		\caption{Performance on diverse edge servers.}
            \vspace{-0.28cm}
		\label{fig:range_length}
\end{minipage}
\begin{minipage}[t]{0.24\textwidth}
\centering
\includegraphics[width=3.7cm]{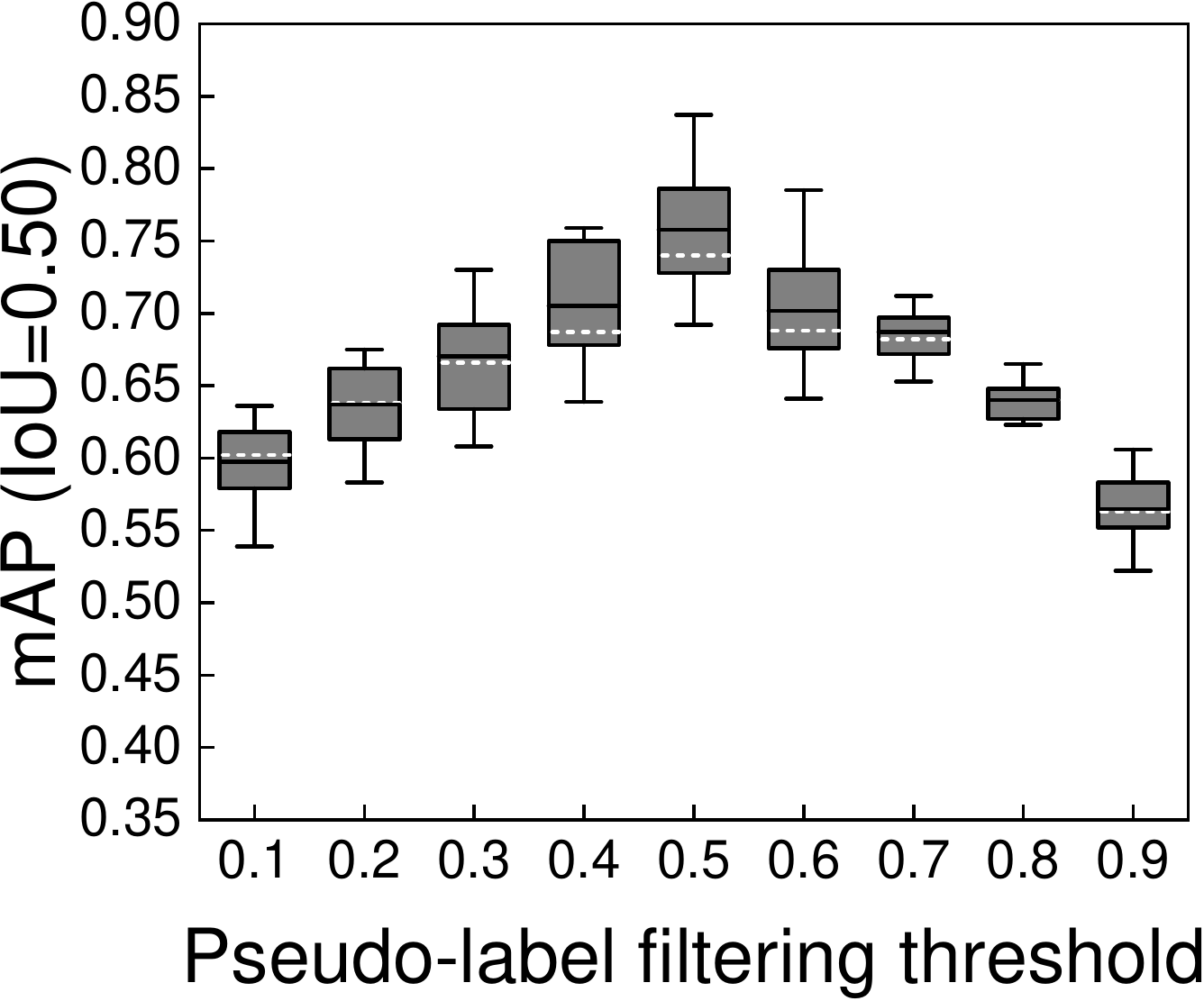}
		\caption{Impact of pseudo-label filtering threshold.}
            \vspace{-0.28cm}
		\label{fig:ex_642}
\end{minipage}
\end{figure}

To demonstrate the impact of the filtering threshold on pseudo-label generation, we conducted a performance comparison under different threshold values, as depicted in Figure~\ref{fig:ex_642}.
As the filtering threshold increased from $0.1$ to $0.5$, the mean Average Precision (mAP) of \sysname increased from $0.597$ to $0.758$. 
This improvement can be attributed to eliminate unreliable pseudo-labels as the threshold increases. 
However, as the filtering threshold increased from $0.5$ to $0.9$, \sysname's mAP gradually decreased. 
This decline can be attributed to the reduction of a significant amount of helpful information, negatively impacting retraining.
Therefore, we have selected $0.5$ as the default filtering threshold. 
\begin{table}[]
\centering
\scriptsize
\caption{Comparison of different grouping strategies.}
\begin{tabular}{|
>{\columncolor[HTML]{FFFFFF}}c |
>{\columncolor[HTML]{FFFFFF}}c 
>{\columncolor[HTML]{FFFFFF}}c 
>{\columncolor[HTML]{FFFFFF}}c |}
\hline
{\color[HTML]{333333} \textbf{Strategy}}     & \multicolumn{3}{c|}{\cellcolor[HTML]{FFFFFF}{\color[HTML]{333333} \textbf{Equal urgency range grouping}}}          \\ \hline
\textbf{Number of groups}          & \multicolumn{1}{c|}{\cellcolor[HTML]{FFFFFF}3}  & \multicolumn{1}{c|}{\cellcolor[HTML]{FFFFFF}4}  & 5  \\ \hline
\textbf{Urgency range length}         & \multicolumn{1}{c|}{\cellcolor[HTML]{FFFFFF}33} & \multicolumn{1}{c|}{\cellcolor[HTML]{FFFFFF}25} & 20 \\ \hline
\textbf{Minimum task probability} & \multicolumn{1}{c|}{\cellcolor[HTML]{FFFFFF}0.24865} & \multicolumn{1}{c|}{\cellcolor[HTML]{FFFFFF}0.1444} & 0.1016 \\ \hline
\textbf{Maximum task probability} & \multicolumn{1}{c|}{\cellcolor[HTML]{FFFFFF}0.5027}  & \multicolumn{1}{c|}{\cellcolor[HTML]{FFFFFF}0.3556} & 0.3286 \\ \hline
{\color[HTML]{333333} \textbf{Strategy}}     & \multicolumn{3}{c|}{\cellcolor[HTML]{FFFFFF}{\color[HTML]{333333} \textbf{Equal task probability grouping}}}        \\ \hline
\textbf{Number of groups}          & \multicolumn{1}{c|}{\cellcolor[HTML]{FFFFFF}3}  & \multicolumn{1}{c|}{\cellcolor[HTML]{FFFFFF}4}  & 5  \\ \hline
\textbf{Task probability}         & \multicolumn{1}{c|}{\cellcolor[HTML]{FFFFFF}0.33}    & \multicolumn{1}{c|}{\cellcolor[HTML]{FFFFFF}0.25}   & 0.2    \\ \hline
\textbf{Minimum urgency range length} & \multicolumn{1}{c|}{\cellcolor[HTML]{FFFFFF}20} & \multicolumn{1}{c|}{\cellcolor[HTML]{FFFFFF}16} & 12 \\ \hline
\textbf{Maximum urgency range length} & \multicolumn{1}{c|}{\cellcolor[HTML]{FFFFFF}40} & \multicolumn{1}{c|}{\cellcolor[HTML]{FFFFFF}34} & 30 \\ \hline
\end{tabular}
\label{tb:echelon}
\end{table}
\begin{table}[t]
\scriptsize
\caption{Error of memory demand profiler.}
\vspace{-0.3cm}
\begin{tabular}{|c|c|c|c|}
\hline
\textbf{Model} & \begin{tabular}[c]{@{}c@{}}\textbf{Memory}\\ \textbf{estimation(MB)}\end{tabular} & \begin{tabular}[c]{@{}c@{}}\textbf{Actual}\\ \textbf{memory(MB)}\end{tabular} & \textbf{Error}  \\ \hline
\textbf{Faster RCNN-ResNet50} & 3718.4 & 3597.0 & 3.38\% \\ \hline
\textbf{Faster RCNN-MobileNetV2} & 3516.8 & 3395.0 & 3.59\% \\ \hline
\textbf{YOLOv3-Darknet53} & 2694.3 & 2577.0 & 4.55\% \\ \hline
\textbf{YOLOv3-ResNet50} & 2554.4 & 2483.0 & 2.88\% \\ \hline
\end{tabular}
\label{tb_651}
\vspace{-0.4cm}

\end{table}
\begin{figure}[t]
  \hfill
  \subfloat[Accuracy gain profiler] {\includegraphics[width=0.21\textwidth]{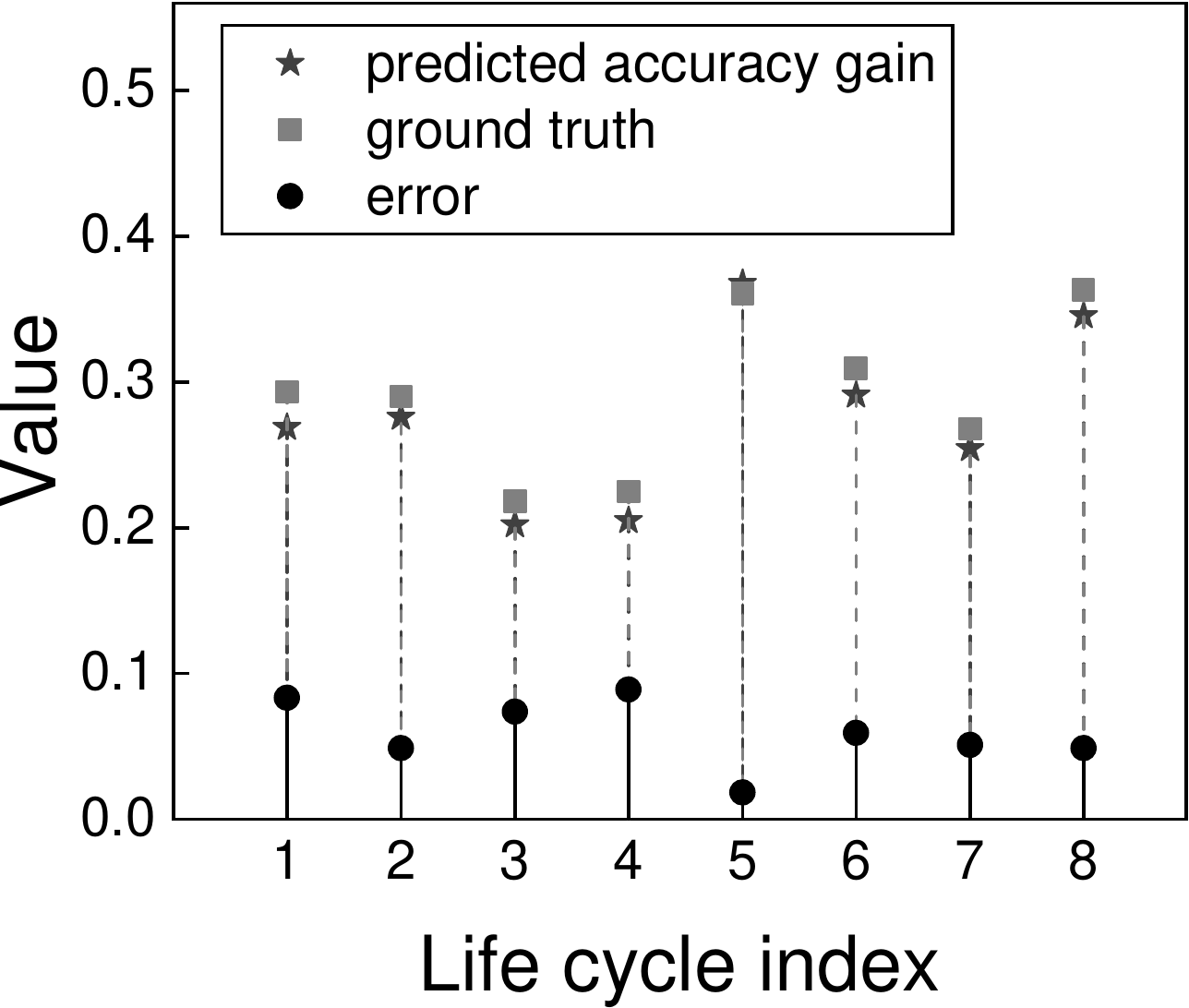}}
  \hfill
  \subfloat[Retraining time profiler] {\includegraphics[width=0.21\textwidth]{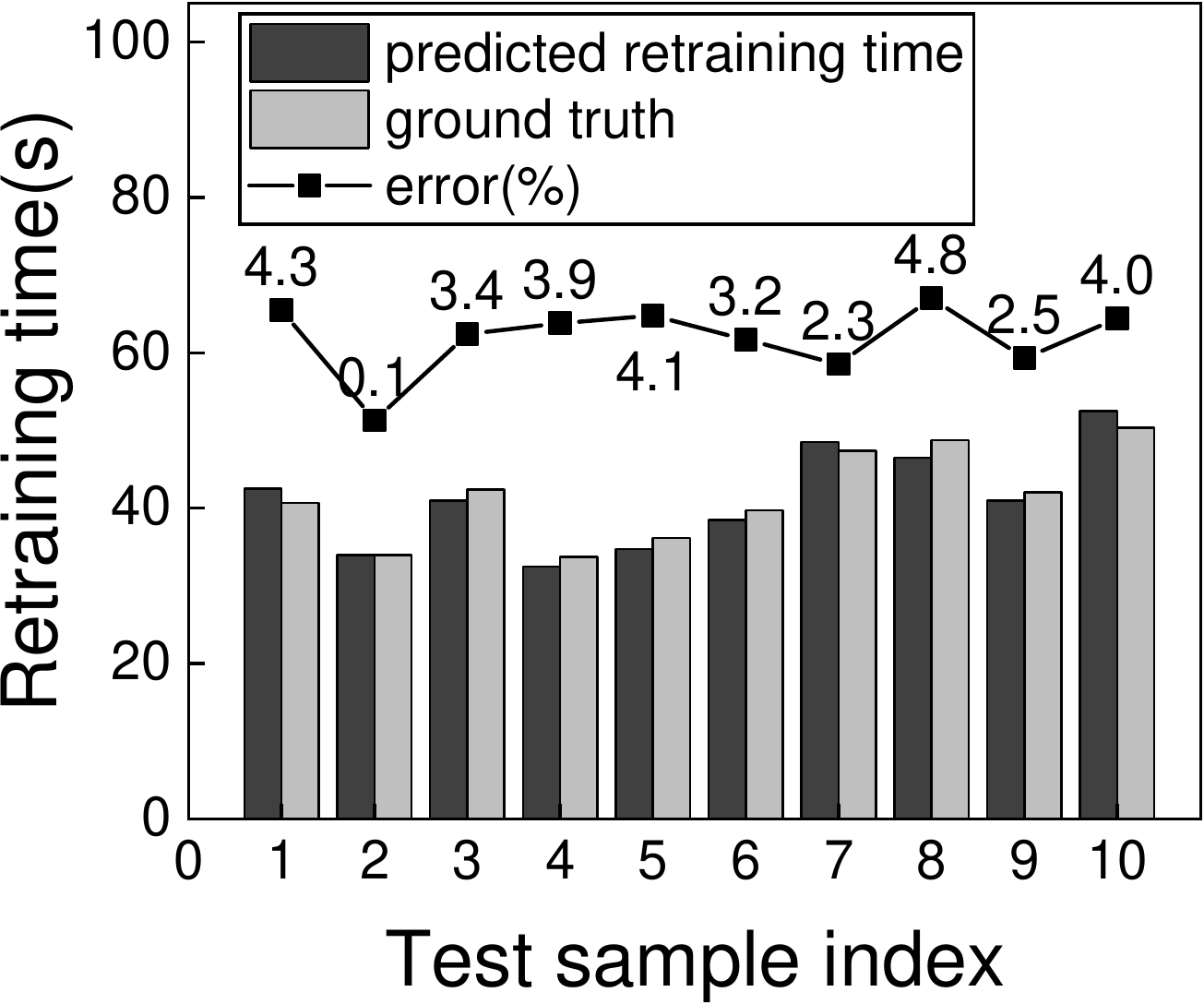}}
  \hspace{0.5cm}
  \caption{Generalization of performance profiler.}

  \label{fig:ex_652_653}
\end{figure}

\subsection{Generalization of Evolution Task Profiler}
\label{subsec:exp_65}

This experiment demonstrates the performance of the proposed mobile DNN evolution task profiler (\S \ref{s_estimator}).
\parahead{Generalization of Memory Demand Profiler}
\label{subsec:exp_651}
Table \ref{tb_651} shows the cross-model performance of the proposed memory demand estimator. 
We use the estimator to predict the retraining memory demands at four different deep models.
The average prediction error is less than $4.55\%$. It is acceptable for the approximate prediction.
\parahead{Generalization of Accuracy Gain Profiler}
\label{subsec:exp_652}
Figure \ref{fig:ex_652_653}(a) shows the accuracy gain estimator's generalization performance across multiple lifecycle evolutions.
Specifically, we adopt different video clips from the testing dataset D1 to test model 2, with which this model experiences eight life cycles and triggers eight model retraining.
The prediction errors between the predicted accuracy gain and ground truth over these life cycles are acceptable, \ie $\leq 5.23\%$.

\parahead{Generalization of Retraining Time Profiler}
\label{subsub_profiler_time}
We run $240$ different evolution tasks to record their retraining time and retraining features (see detail features in \S \ref{s_estimator}).
The SGD optimizer~\cite{bib:sgd:Robbins} with a learning rate of $0.0005$ and a momentum of $0.9$ is used to train the estimation network by $5000$ epochs. 
We use $200$ records for estimation network training and $40$ records for generalization performance testing.
Figure \ref{fig:ex_652_653}(b) shows the predicted retraining time, ground truth, and corresponding errors for ten randomly selected video clips from D2. 
The prediction error is $\leq 4.8\%$.
Besides, we run this estimation network $1000$ times to test its overhead in prediction latency, which is negligible ($\leq 0.09ms$).

\section{Related Work}
\label{sec:related}
In this section, we discuss the closely related works.

\parahead{On-device Mobile Deep Vision Applications}
On-device inference using deep neural network~\cite{bib:Computer17:Ananthanarayanan, bib:mobicom2018:fang, bib:osdi2018:hsieh} 
is the key enabler for diverse mobile applications,
including mobile VR/AR~\cite{bib:mmsys2019:shi}, autonomous human-following drones~\cite{bib:itoit2018:ke}, 
vision-based robot navigation~\cite{bib:icufn2018:kim}, and autonomous driving cars~\cite{bib:iccv2015:chen}.
%
%
%
Enormous DNN compression techniques~\cite{bib:iclr2016:han,bib:cvpr2019:wang,bib:nips2015:han,bib:iclr2016:li} 
have been proposed to facilitate the deployment of compute-intensive DNN on resource-constrained mobile ends
by reducing the model complexity~\cite{bib:nips2015:han, bib:cvpr2019:kim,bib:nips2015:novikov, bib:arXiv2015:Hinton,bib:aaai2019:heo}.
%
\sysname follows this trend of adopting compressed deep models for enabling on-device mobile vision applications.

\parahead{Continuous DNN Evolution to Adapt to Data Drift}
\textit{Data drift} refers to changes between new data and source datasets in joint distribution, which violates the IID assumption~\cite{bib:quinonero2008dataset, bib:moreno2012unifying}.
Data drift always leads to accuracy degradation~\cite{bib:gama2014survey}, which worsens the compressed DNN with insufficient generalization~\cite{bib:nsdi2022:Bhardwaj}.
%
Many previous efforts have exploited to tackle data drift, mainly from three complementary perspectives, \ie continuous retraining~\cite{bib:nsdi2022:Bhardwaj,bib:IITJ2022_Jia,bib:cvpr22_tiezzi},meta-learning~\cite{bib:han2022}, domain adaptation~\cite{bib:zhang2014domain,bib:zara2023}, and online knowledge distillation~\cite{bib:iccv2019:Mullapudi, bib:iccv2021:Khani, bib:icml2021:zhu}.
These methods handle data drift by periodically evolving models on new data and differ in learning goals and strategies.
However, they fail to analyze the data drift comprehensively, so they select retraining data one-sidedly~\cite{bib:iccv2021:Khani,bib:iccv2019:Mullapudi,bib:IITJ2022_Jia} or even adopt a fixed sampling rate~\cite{bib:iccv2021:Khani}, leading to the loss of key data and causing unsatisfactory retraining effects.
In addition, they aim to optimize the inference accuracy for a single compressed DNN. Still, they fail to improve balancing competitions among multiple requests from mobile ends for asynchronous retraining in a system.
\sysname advances them in the following aspects: \textit{(i)} It balances the competition between numerous asynchronous evolution requests, thereby improving the average QoE of all users. \textit{(ii)} It analyzes different types of data drifts at the mobile end and then designs the corresponding evolution scheme.

\parahead{Edge-assisted Mobile Systems for Live Video Analytics}
Recent mobile systems also try to fully utilize the computational resources 
inside camera-embedded mobile ends and other edge devices for \textit{live video} analytics.
For example, 
Reducto\cite{bib:sigcomm2020:li} implements on-camera filtering for real-time video analysis.
DDS\cite{bib:sigcomm2020:du} improves the video analytics system throughput by adaptively balancing the workloads across cameras.
%
ELF\cite{bib:mobicom2021:Zhang} accelerates mobile vision applications by parallelly offloading computation to edge clusters. 
Glimpse\cite{bib:sensys2015:chen} splits the computation between mobile and server devices to enable real-time object recognition.
%
\sysname is built upon this thread of efforts.
It conducts live video analytics on mobile ends and offloads the retraining to the edge server, continually optimizing the accuracy of multiple mobile vision applications.

\section{Conclusion}
\label{sec:conclude}
This paper presents \sysname, 
a framework that maximizes the performance of edge-assisted continuous model evolution systems.
\sysname detects the type of data drift in the mobile end to quantitatively predicts the accuracy drop on live videos and proposes the data drift-aware video frame sampling method for balancing the size of uploaded live video and the accuracy of the evolved deep models.
In addition, \sysname fairly shares the resources of the edge server among multiple mobile ends and simultaneously maximizes the average QoE of all mobile ends,
by allocating the limited computing and memory resources on the edge server and the competition between asynchronous evolution tasks initiated by different mobile ends.

\bibliographystyle{IEEEtran}


\end{document}